\documentclass[twocolumn,showpacs,superscriptaddress,floatfix]{revtex4}

\usepackage{graphicx}
\usepackage{dcolumn}
\usepackage{bm}
\usepackage{color,stmaryrd}
\usepackage[normalem]{ulem}
\usepackage{mathrsfs}
\usepackage{amsmath}
\usepackage{amssymb}
\usepackage{times}

\newcommand*{\img}[1]{%
\raisebox{-.3\baselineskip}{%
\includegraphics[
height=\baselineskip,
width=\baselineskip,
trim={6.5cm 8cm 6.5cm 9cm}, clip, 
]{#1}%
}%
}

\newcommand*{\imgl}[1]{%
\raisebox{-.3\baselineskip}{%
\includegraphics[
height=\baselineskip,
width=\baselineskip,
trim={7.5cm 8cm 7.5cm 8cm}, clip, 
]{#1}%
}%
}

\newcommand*{\imga}[1]{%
\raisebox{-.3\baselineskip}{%
\includegraphics[
height=\baselineskip,
width=\baselineskip,
trim={8.0cm 10cm 8.0cm 10cm}, clip, 
]{#1}%
}%
}

\begin{document}

\title{Adding color: Visualization of energy landscapes in spin glasses}

\author{Katja Biswas}
\affiliation{Department of Physics and Astronomy, Texas A\&M University, College Station, Texas 77843-4242, USA}

\author{Helmut G. Katzgraber}
\affiliation{Microsoft Quantum, Redmond, Washington 98052, USA}

\begin{abstract}

Disconnectivity graphs are used to visualize the minima and the lowest
energy barriers between the minima of complex systems. They give an easy
and intuitive understanding of the underlying energy landscape and, as
such, are excellent tools for understanding the complexity involved in
finding low-lying or global minima of such systems.  We have developed a
classification scheme that categorizes highly-degenerate minima of spin
glasses based on similarity and accessibility of the individual states.
This classification allows us to condense the information pertained in
different dales of the energy landscape to a single representation using
color to distinguish its type and a bar chart to indicate the average
size of the dales at their respective energy levels. We use this
classification to visualize disconnectivity graphs of small
representations of different tile-planted models of spin glasses.  An
analysis of the results shows that different models have distinctly
different features in the total number of minima, the distribution of
the minima with respect to the ground state, the barrier height and in
the occurrence of the different types of minimum energy dales.

\end{abstract}


\maketitle

\section{Introduction}

Originally introduced to describe energy landscapes of
tetrapeptides~\cite{becker:97}, disconnectivity graphs have become a
powerful tool in visualizing the complex relationships between the
different energy minima of bio-polymers and
nano-cluster~\cite{brooks:01,wales:04,wales:05}. Disconnectivity graphs
represent a complex multi-dimensional energy landscape via a set of
minima and the connections between them. The connections illustrate the
lowest energy barrier a system has to overcome to be able to transition
between different minima. In this respect, the disconnectivity graphs
give a clear visual aid and may also serve as a tool in understanding
some of the difficulties when dealing with the optimization of complex
systems and processes. Moreover, with the insights gained, bespoke
nontrivial random benchmark problems can be designed to benchmark
optimization tools both classical
\cite{wang:13b,isakov:15,tsukamoto:17,matsubara:17,aramon:19,hamerly:19}
and quantum
\cite{ronnow:14a,heim:15,hen:15a,mandra:16b,mandra:17a,mandra:18}.

In spin glasses complexity originates from the interplay of frustration
and disorder \cite{toulouse:77,kirkpatrick:77,anderson:78}, and can have
an effect on the dynamics and success rates of optimization routines.
Trying to gain an understanding of the underlying energy structures of
spin glasses has a recurring history. Garstecki {\em et
al.}~\cite{garstecki:99} used enumeration to draw disconnectivity graphs
and dynamical connectivity graphs for small two-dimensional Ising spin
systems.  Amoruso {\em et al.}~\cite{amoruso:06} used a hierarchical
approach to calculate minimum energy barriers. Burda {\em et al.}~used
the lid algorithm for the barriers~\cite{burda:06} and a steepest
descent method to find internal structures~\cite{burda:07}. The branch
and bound algorithm was used to gain information about the energy
landscape for spin systems up to the third
excitation~\cite{krawczyk:02}. Dall and Sibani~\cite{dall:03} used aging
dynamics to gain some insight into the structure of valleys and barriers
of the energy landscape of Ising spin glasses. Barrier
trees~\cite{fontanari:02,hordijk:03} have been used to describe $p$-spin
models. Seyed-Allaei {\em et al.}~\cite{seyed-allaei:08} used
disconnectivity graphs to study the energy-landscape of up to $27$ spin
Ising models. Finally, Zhou {\em et al.}~\cite{zhou:09,zhou:11} utilized
random walks to obtain minima and barriers and thus produce
disconnectivity graphs.

Minimum-energy configurations that have regions of spins for which the
local Hamiltonian is zero, broaden the minima and form dales (wide
valleys) of equal energy in the energy landscape. For larger systems,
these dales~\cite{comment:name} add difficulties to the visualization of
disconnectivity graphs as they essentially consist of a vast number of
minima connected via their own energy and thus do add very little to an
intuitive understanding of the underlying energy structure of the
system. For this reason we propose to simplify the visualization of
these structures by merging them into representations that we refer to
as ``dale minima.'' We further distinguish between different types of
dale minima, which are then visualized via colors in the disconnectivity
graph, and indicate their respective sizes by bar charts for the
different energy levels. This greatly enhances the intuitive
understanding of the underlying energy landscapes and simplifies the
representations, allowing for the mapping of larger structures and
higher energies while preserving the core information of the landscape.
The idea of grouping minima has been utilized in the ballistic search
method \cite{hartmann:00a,hartmann:00b,hartmann:02,mann:10} to find
ground-state configurations for various problems. In ballistic search,
minima that are connected by a single variable or single spin flip at no
energy cost are grouped into one general category which in
Refs.~\cite{hartmann:00a,hartmann:00b,hartmann:02,mann:10} are called
``clusters.'' However, our approach is different in that we further
distinguish between the different ways the minima can be connected.

A different approach to obtain information about ground states was given
by Landry and Coppersmith \cite{landry:02} who define a cluster as
groups of minima connected in a way analogous to the ballistic search
method, but further distinguish between different structures that can
occur within a cluster which they call a ``bunch.'' In their definition,
a bunch is a local structure of spins that are either zero-energy spins
or under certain fulfilled conditions can become zero-energy spins. In
this way, a single configuration can have multiple bunches, which are
then used to obtain information about the structure of the individual
ground states. Our approach also differs from the aforementioned bunches
in that, to ease the visualization, we use the overall connectivity
between the individual minima. In other words, although the connection
between minima is influenced by local structures of spins within the
particular minimum energy configurations, we are not interested in
characterizing these local structures. We characterize the entire
minimum to belong to a certain type which we call either a regular, or a
dale minimum. This approach allows us to effectively enhance and reduce
the visualization of minima in disconnectivity graphs, which is the main
goal of this paper.

To demonstrate the visualization approach we propose in this work, we
use our enhanced disconnectivity graphs to study the underlying energy
landscape of instances of planar tile-planted spin
glasses~\cite{hamze:18,perera:19}. These are special types of tunable
spin glasses used to benchmark optimization techniques that have been
constructed with a known ground-state solution. Note that other types of
Ising Hamiltonians with planted solutions have been constructed on
Chimera
graphs~\cite{hen:15a,martin-mayor:15,katzgraber:14,katzgraber:15}, using
a random adaptive optimization method~\cite{marshall:16}, as well as via
patch planting~\cite{wang:17}.

The article is structured as follows. In section~\ref{sec:classi} we
introduce and explain our classification scheme.
Section~\ref{sec:discon} explains the basics of disconnectivity graph.
Section~\ref{sec:model} gives a short overview of the tile-planted spin
glasses constructed with different elementary plaquettes and which we
refer to as $C_1$-type model, $C_2$-type model, and $C_3$-type model. In
section~\ref{sec:results} we discuss our results, followed by
conclusions.

\section{Methods}

\subsection{Classification}
\label{sec:classi}

Disconnectivity graphs are two-dimensional representations of
high-dimensional potential energy landscapes. In their simplest form,
they depict the minima of the landscape and the lowest high-energy
barriers between any two minima. The lowest high-energy barrier is the
minimum increase in energy necessary to transition from one minimum to
another.  When dealing with disordered spin systems several
factors are to be taken into account that complicate the generation of
disconnectivity graphs.

For big enough systems of spin-glasses in which the spin-spin interactions 
are drawn from a discrete and finite distribution, there exist
arrangements of the spin-spin interaction that allow some of the spins
to take any orientation without changing the energy of the system. This
is typically the case when the local interactions centered around a spin
$s_i$
and their corresponding connecting spins sum to zero, i.e. $H_{E_i}=0$.
 We call the spins $s_i$ 
which show this effect zero-energy spins, thus highlighting the
fact that their contribution to the total energy of the system is zero.
These effects can influence isolated spins or groups of connected
spins. Furthermore, certain arrangements of spins and interactions can
``transfer'' the effects of zero local energy to other spins, thus leading to
pathways in the energy landscape that can only be traversed in specific
directions {\em without increasing the energy}. A different orientation
of a single spin constitutes a different configuration of the whole
system. Hence, configurations that only differ by spin orientations of
the described form constitute dales in the energy landscape. A dale here
is a region on the energy landscape that has the same energy for
different configurations of the system \cite{comment:hi}, i.e., the
configurations are linked to each other by paths in the energy landscape
that do not require an intermediate increase in energy. From a practical
point of view when generating disconnectivity graphs, the minima
belonging to the same dale increase the storage of data substantially
while adding very little information to the energy structure of the
system. In addition, dales clutter disconnectivity graphs with
unnecessary information.  From an energy connectivity perspective, one
can think of each such a dale as an individual but degenerate state.
This has led us to develop a color scheme for the disconnectivity graphs
to distinguish between regular (nondegenerate) minima and various types
of dales.

We distinguish between different types of minima. The
types are based on similarity arguments and accessibility of the minima
within the dales. We distinguish the minima as follows:\\ \\
\textbf{Regular minimum}: A regular minimum is a minimum for which the 
flip of any spin of the system increases the energy.\\ \\
\textbf {Type-1 dale minimum}: A dale minimum of type-1 is a degenerate 
minimum in which any combination of flips of the zero-energy spins 
preserves the energy of the system.\\ \\
\textbf{Type-2 dale minimum}: A dale minimum of type-2 is a degenerate 
minimum in which only some, but not all, combinations of flips of 
the zero-energy spins preserve the energy of the system.\\ \\
\textbf{Type-3 dale minimum}: A dale minimum of type-3 represents a 
connection of dale minima accessible only via specific 
transitions between the minima.

We further separate type-3 dale minima into two subcategories, which we
call ``type-3 dale minima with a simple path'' and ``type-3 dale minima
with a split path.'' These subcategories are based on the specific ways
the energy landscape has to be traversed in order to preserve the energy
of the system.  Furthermore, note that the classification into regular
minima, type-1 dale minima and type-2 dale minima is based on the
overall connectivity of the individual configurations of the whole
configuration space, i.e., while the zero-energy spins can form multiple
small clusters within the systems, the classification according to the
types is determined by the occurrence of the structures of connecting
spins of the highest type. This is justified by the fact that these
individual structures are the ones that determine the connectivity of
the individual configurations on the configuration space. We also
emphasize that the present study is performed for sparse graphs with
fixed connectivity. A different, more complex, picture might emerge for
Ising Hamiltonians on random, as well as fully-connected graphs.

In the disconnectivity graphs, we distinguish the different types of
minima using colors. The regular minima are depicted in black, whereas
type-1 and type-2 dale minima are colored in blue and green,
respectively.  Because type-3 dale minima are two dales joined together,
we represent them by connecting the corresponding dales by a horizontal
line (colored in red) at their respective energies indicating the
zero-energy transition between the two connecting dales.

\begin{figure}
\includegraphics[width=0.45\linewidth,trim={0 2cm 0 0},clip]{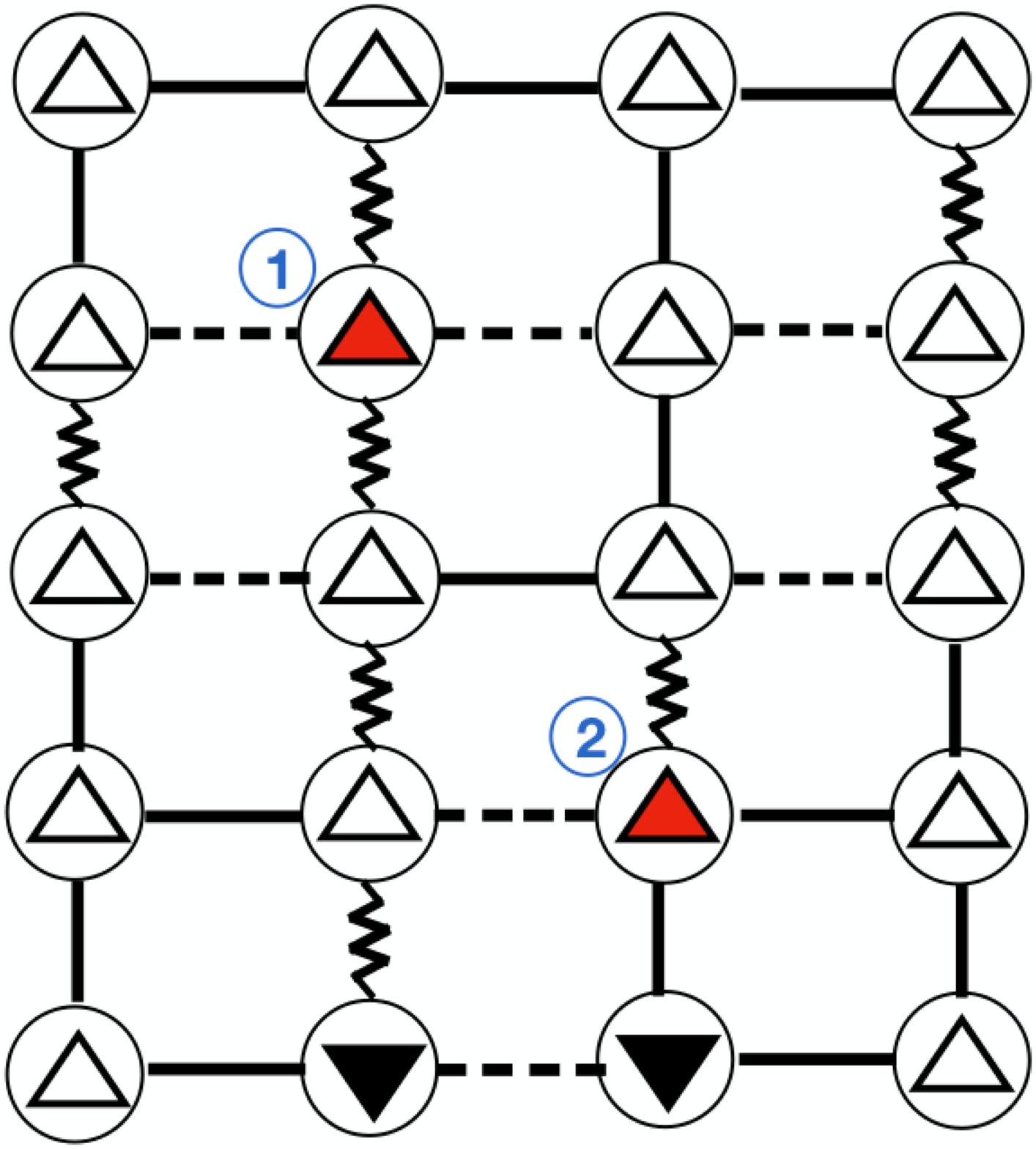}
\caption{\label{pseudo1}
Visualization of type-1 dale minima. The interactions $J_{ij} = -1$,
$1$, and $2$ are represented by wiggly lines (\protect\imgl{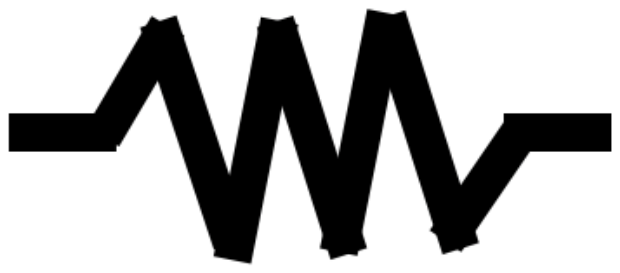}),
dashed lines (\protect\imgl{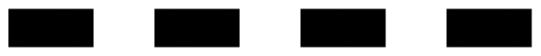}), and solid lines
(\protect\imgl{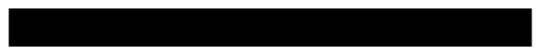}), respectively. White triangles
\protect\img{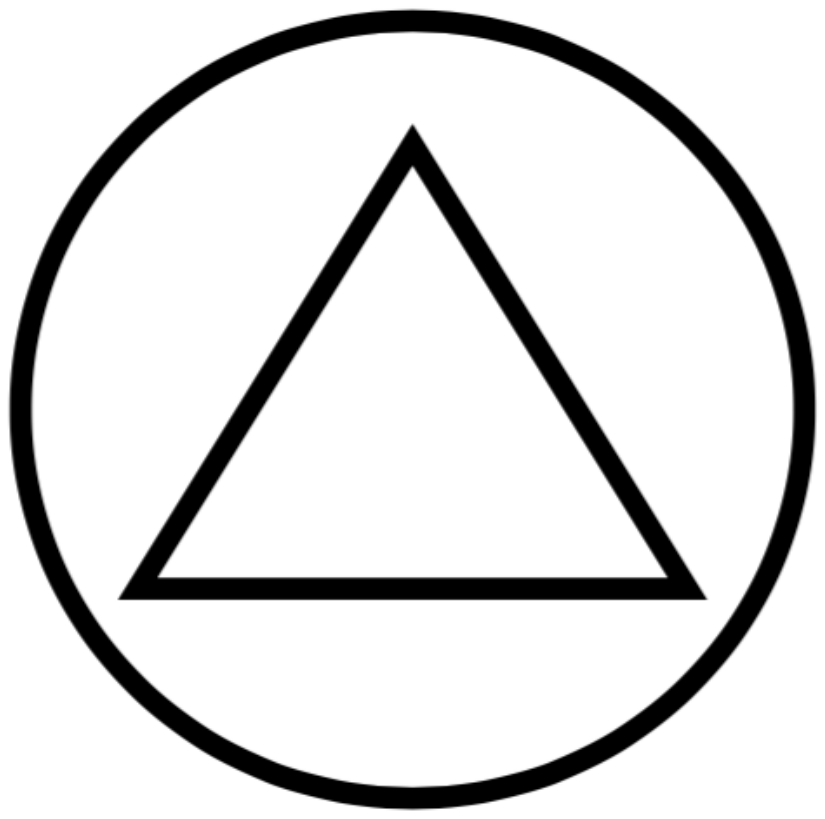} represent spin up and black triangles
\protect\img{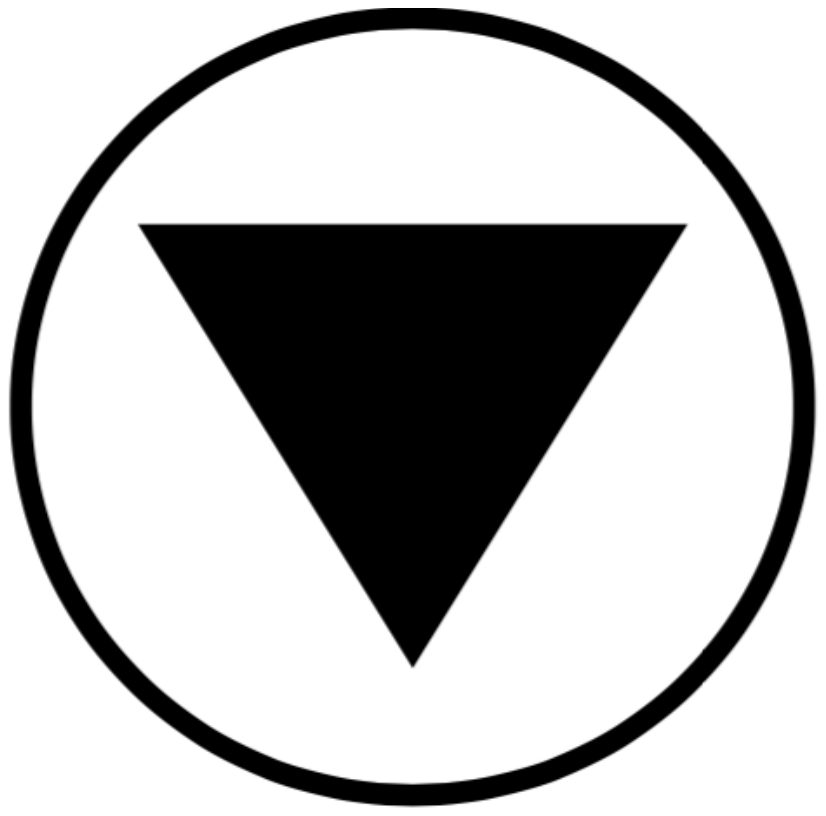} represent spin down. The red triangles
\protect\img{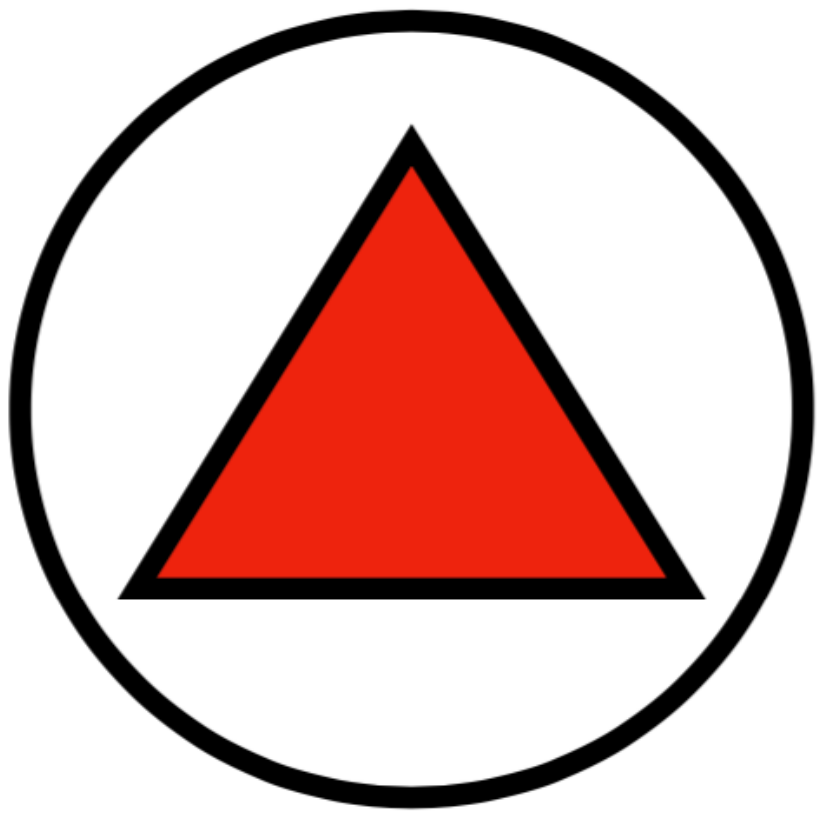} labeled \protect\img{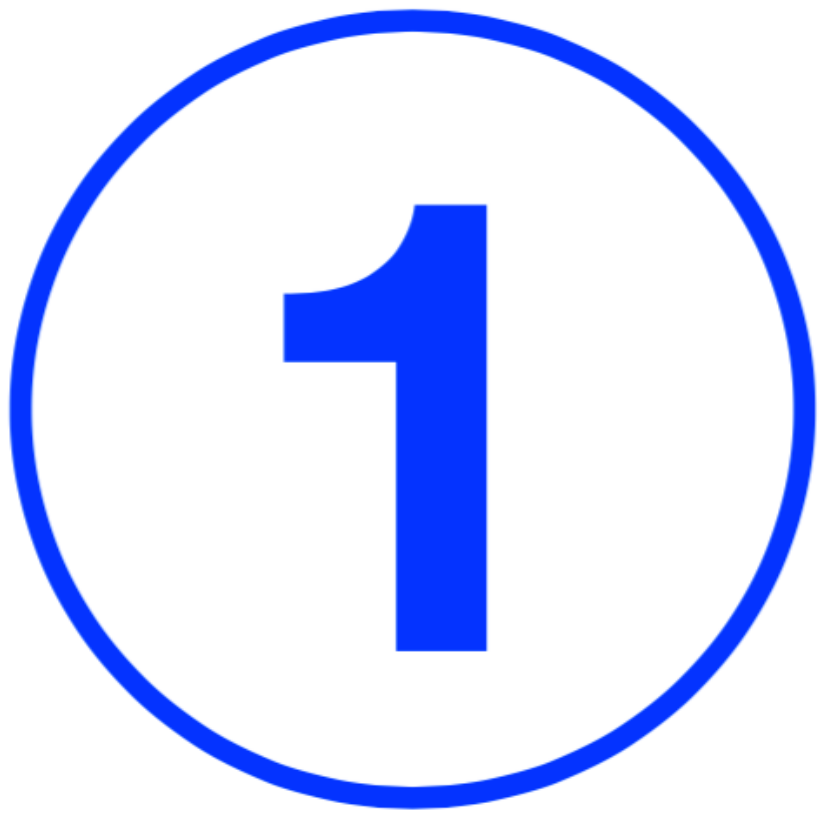} and
\protect\img{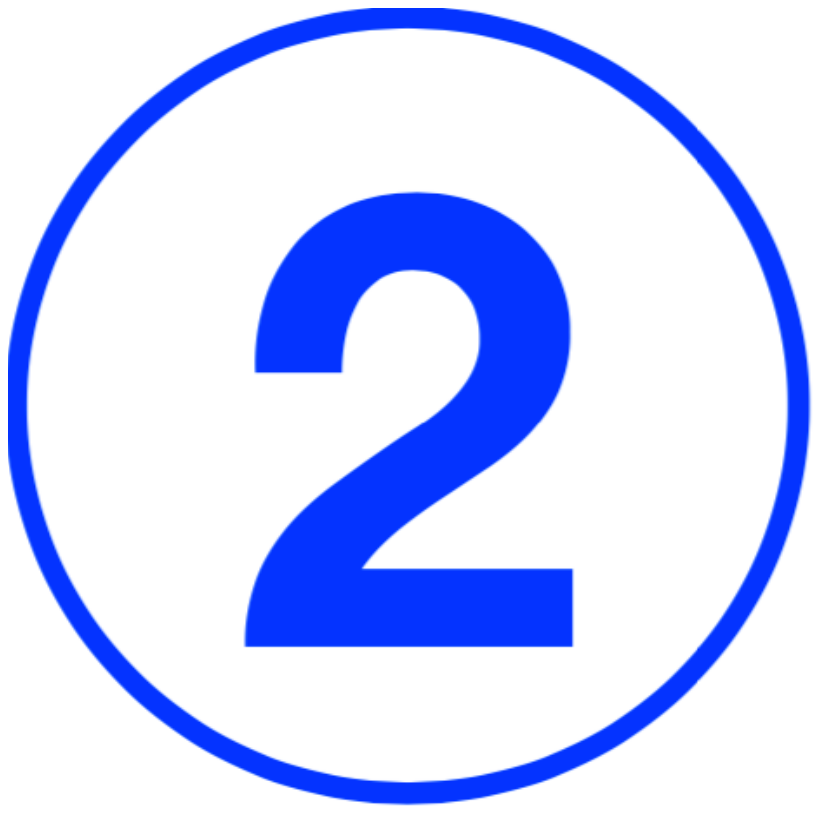} are zero-energy spins. These can be flipped in
any combination at no energy cost.}
\end{figure}

\begin{figure}
\includegraphics[width=0.45\linewidth,trim={2cm 5cm 0 0},clip]{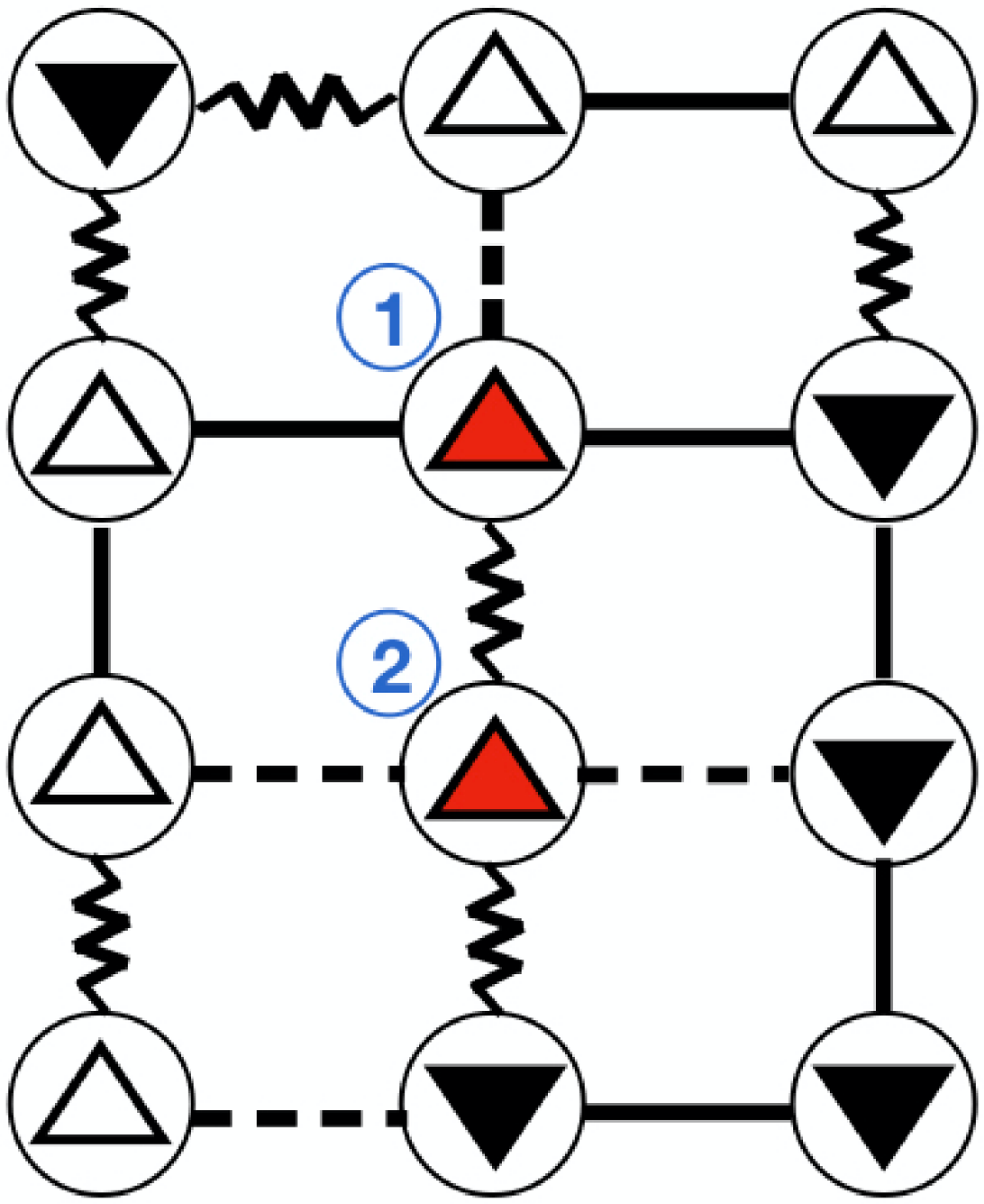}
\caption{\label{pseudo2}
Visualization of type-2 dale minima. The interactions $J_{ij} = -1$,
$1$, and $2$ are represented by wiggly lines (\protect\imgl{fig01}),
dashed lines (\protect\imgl{fig02}), and solid lines
(\protect\imgl{fig03}), respectively. White triangles
\protect\img{fig04} represent spin up and black triangles
\protect\img{fig05} represent spin down. The red triangles
\protect\img{fig06} labeled \protect\img{fig07} and
\protect\img{fig08} are zero-energy spins. They can be flipped
only in certain combinations at no energy cost.}
\end{figure}

\begin{figure}
\includegraphics[width=0.5\linewidth,  trim={0 5cm 0 0}, clip]{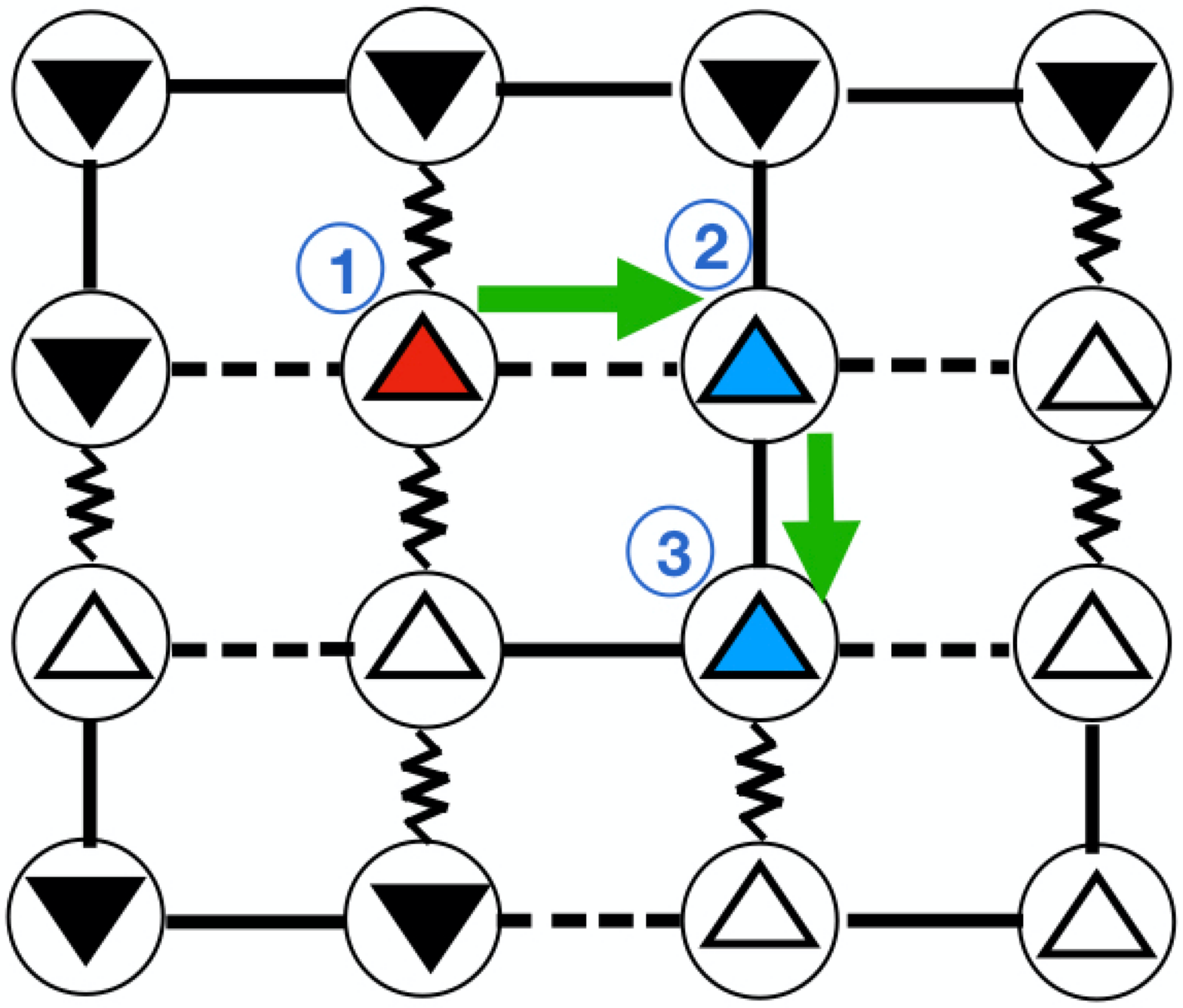}
\caption{\label{pseudo3_1}
Visualization of type-3 dale minima with a simple path. The interactions
$J_{ij} = -1$, $1$, and $2$ are represented by wiggly lines
(\protect\imgl{fig01}), dashed lines (\protect\imgl{fig02}), and
solid lines (\protect\imgl{fig03}), respectively. White triangles
\protect\img{fig04} represent spin up and black triangles
\protect\img{fig05} represent spin down. In the presented
configuration only the red triangle \protect\img{fig06}
labeled \protect\img{fig07} is a zero-energy spin and can be
flipped at no energy cost. However, once spin \protect\img{fig07}
has been flipped, the blue spin \protect\img{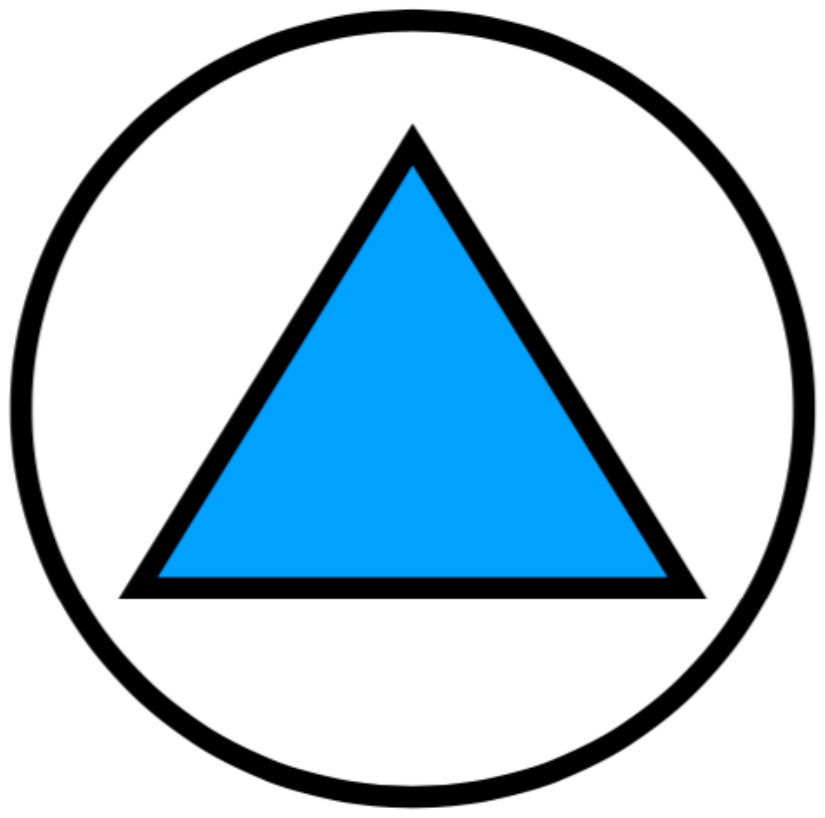} labeled
as \protect\img{fig08} becomes a zero-energy spin.  Lastly, only
after \protect\img{fig08} flips the spin labeled
\protect\img{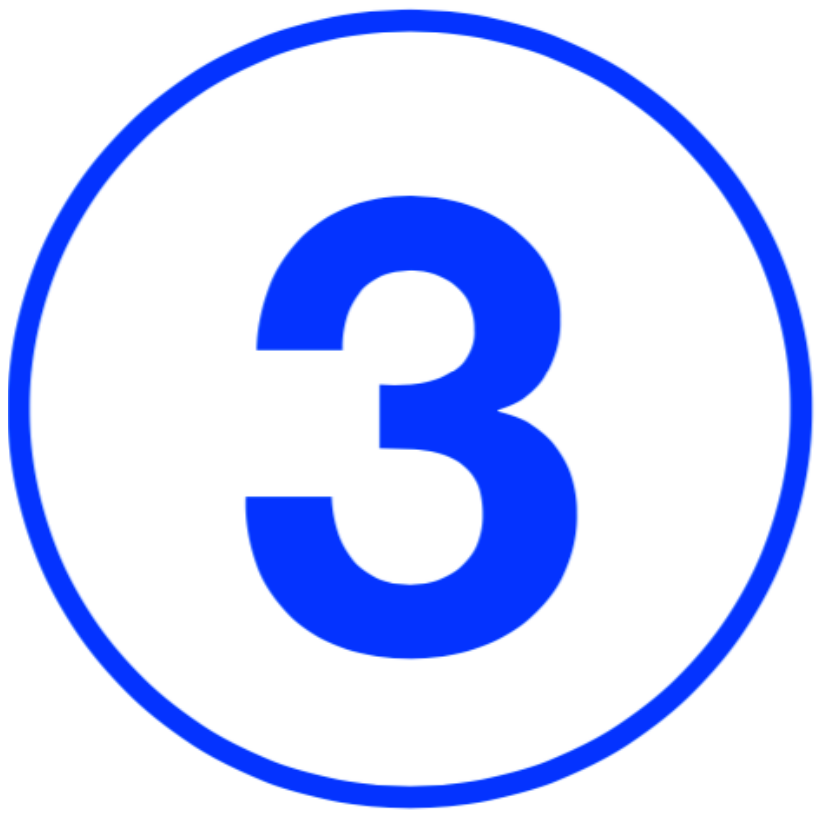} can be flipped at no energy cost. The
direction of this simple path is indicated by the green arrows
(\protect\imga{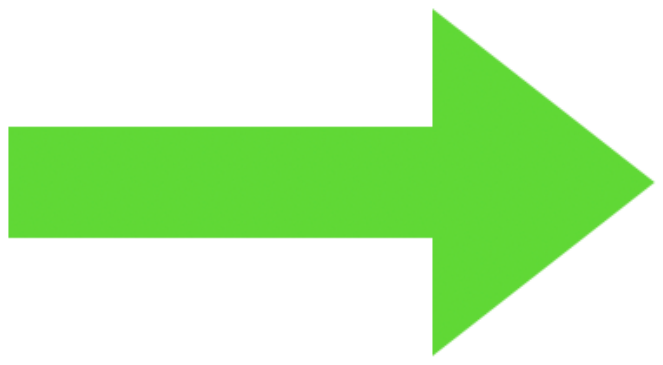}). }
\end{figure}

\begin{figure}
\includegraphics[width=0.45\linewidth, angle=270, trim={0 2cm 2cm 0}, clip]{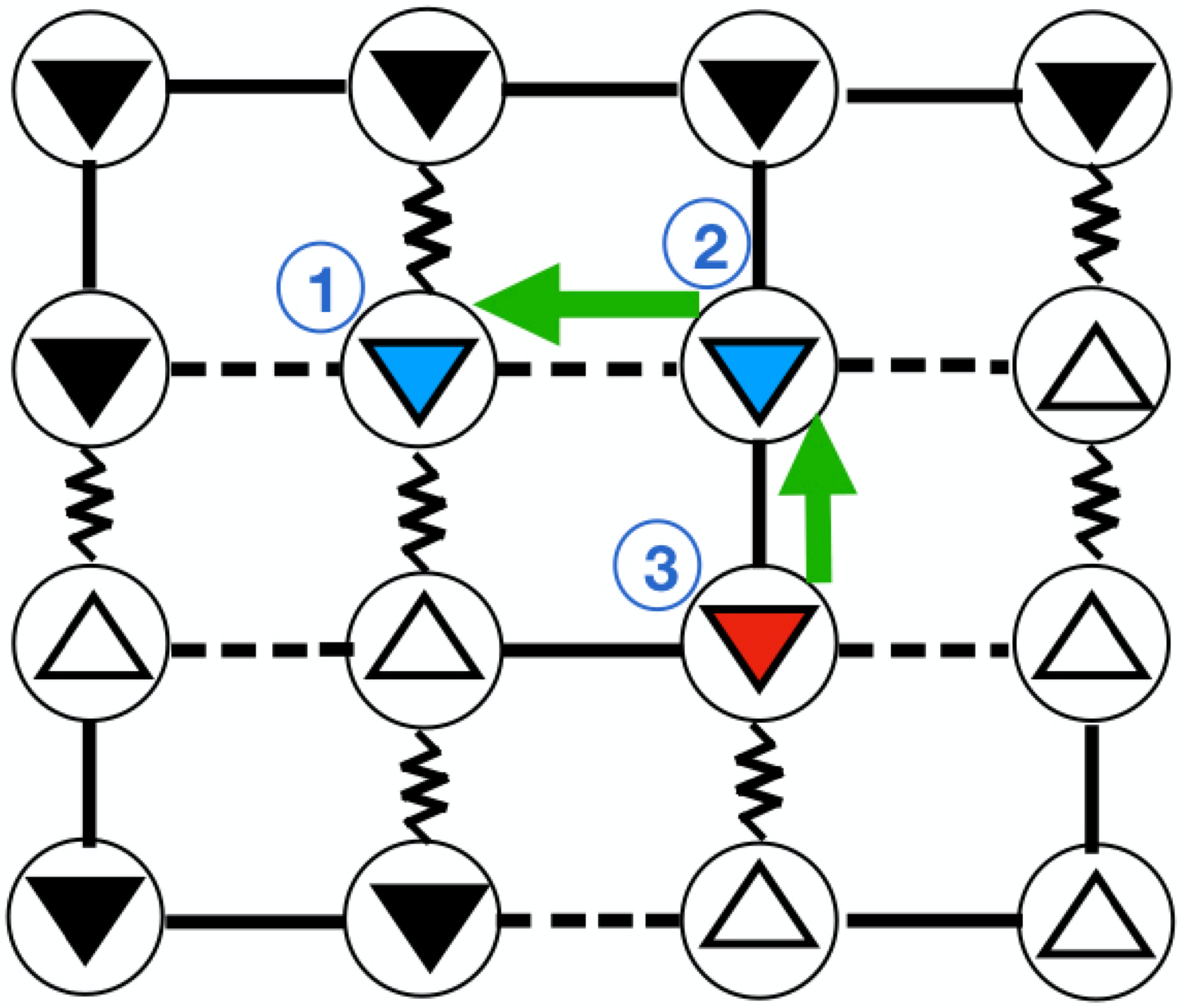}
\caption{
\label{pseudo3_2}Visualization of type-3 dale minima with a simple path
in the reverse direction of Fig.~\ref{pseudo3_1}.  The interactions
$J_{ij} = -1$, $1$, and $2$ are represented by wiggly lines
(\protect\imgl{fig01}), dashed lines (\protect\imgl{fig02}), and
solid lines (\protect\imgl{fig03}), respectively. White triangles
\protect\img{fig04} represent spin up and black triangles
\protect\img{fig05} represent spin down.  In the presented
configuration only the red triangle \protect\img{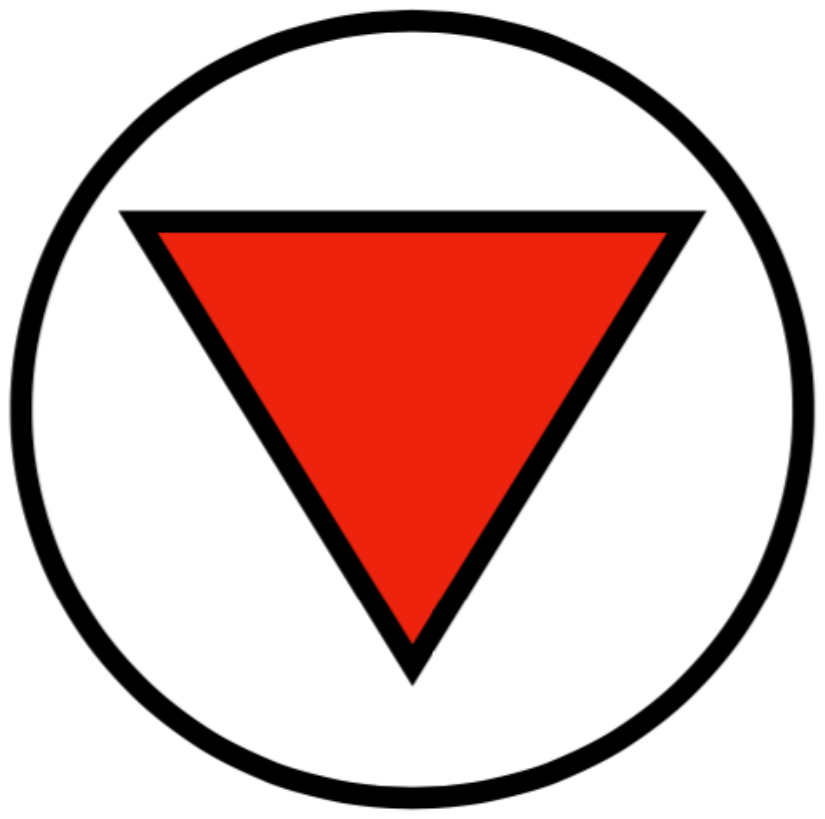}
labeled \protect\img{fig010} is a zero-energy spin and can be
flipped at no energy cost. However, once spin
\protect\img{fig010} has been flipped the blue spin
\protect\img{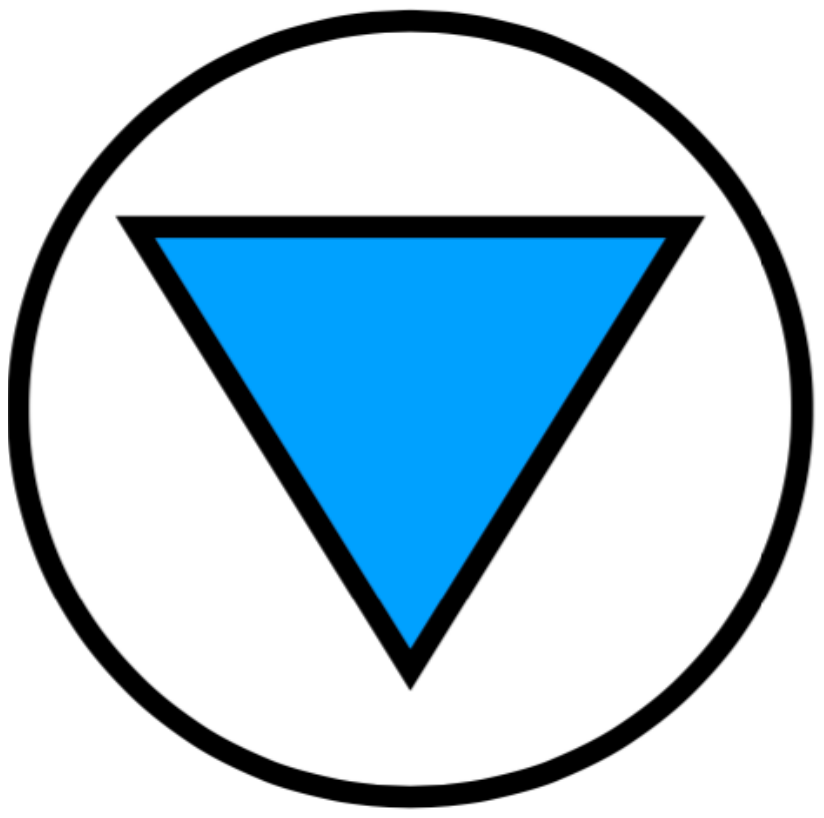} labeled with \protect\img{fig08}
becomes a zero-energy spin. And lastly, only after
\protect\img{fig08} flips the spin labeled
\protect\img{fig07} can be flipped at no energy cost. The
direction of this simple path is indicated by the green arrows
(\protect\imga{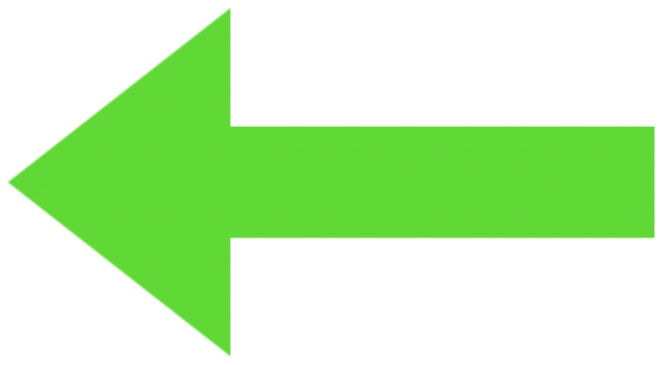}). Note that this path is in the
reverse direction of Fig.~\ref{pseudo3_1}. }
\end{figure}

\begin{figure}
\includegraphics[width=0.65\linewidth, angle=270, trim={0 5cm 0 0}, clip]{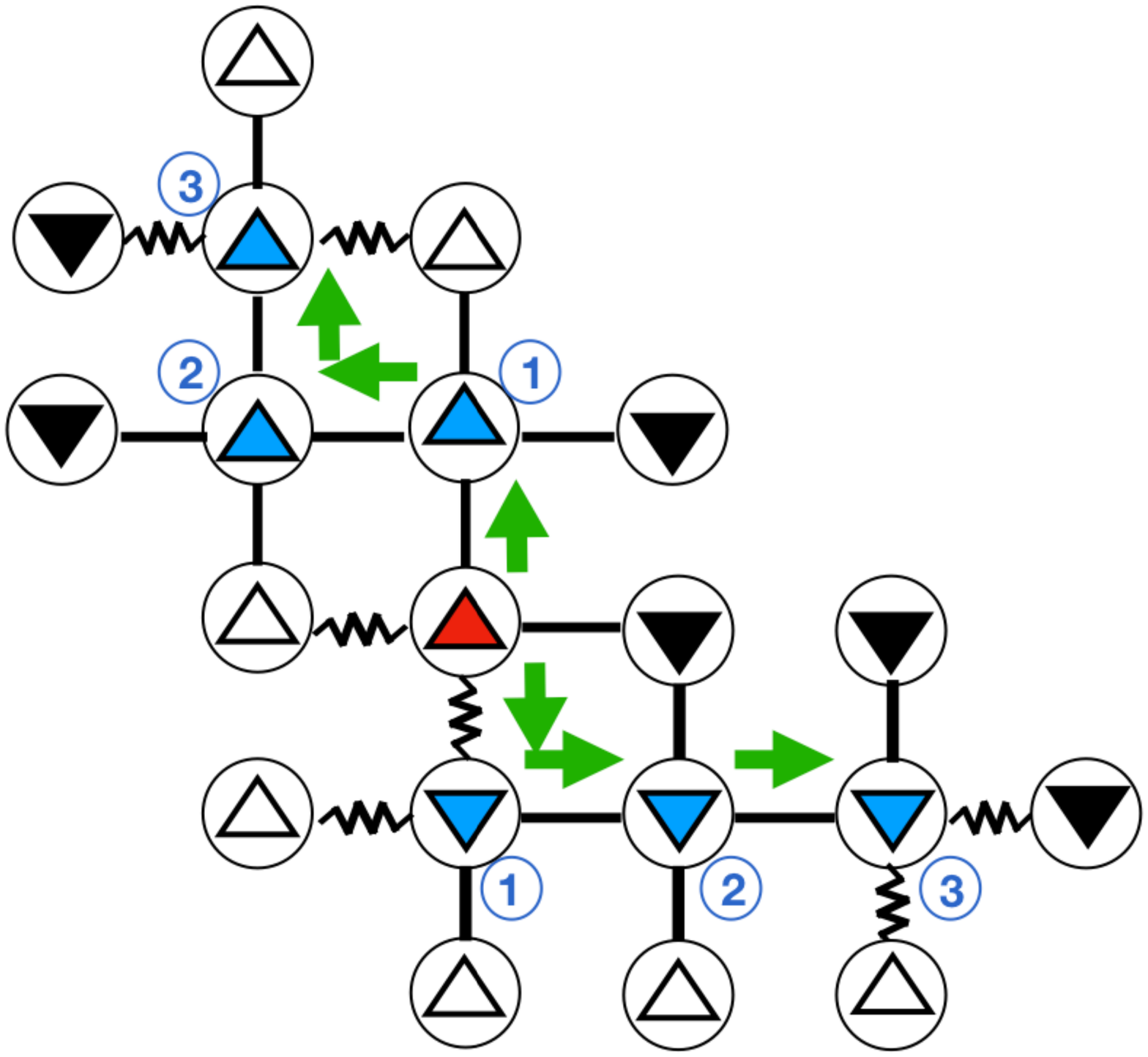}
\caption{ \label{pseudo3_sub2_1}
Visualization of type-3 dale minima with a split path.  The interactions
$J_{ij} = -1$, $1$, and $2$ are represented by wiggly lines
(\protect\imgl{fig01}), dashed lines (\protect\imgl{fig02}), and
solid lines (\protect\imgl{fig03}), respectively. White triangles
\protect\img{fig04} represent spin up and black triangles
\protect\img{fig05} represent spin down.  In the configuration
shown, only the red triangle \protect\img{fig06} is a
zero-energy spin and can be flipped at no energy cost. Once the
zero-energy spin has been flipped the spins drawn in blue
\protect\img{fig09} \protect\img{fig012} which
are labeled \protect\img{fig07} become zero-energy spins, and if
\protect\img{fig07} have been flipped, the blue spins labeled
\protect\img{fig08} can be flipped at no energy cost. This then
allows to flip spin \protect\img{fig010}. Starting with flipping
the center spin \protect\img{fig06}, this dale splits into two
paths, indicated by the green arrows \protect\imga{fig011}.}
\end{figure}

\begin{figure}
\includegraphics[width=0.65\linewidth, angle = 270,  trim={0 5cm 0 0}, clip]{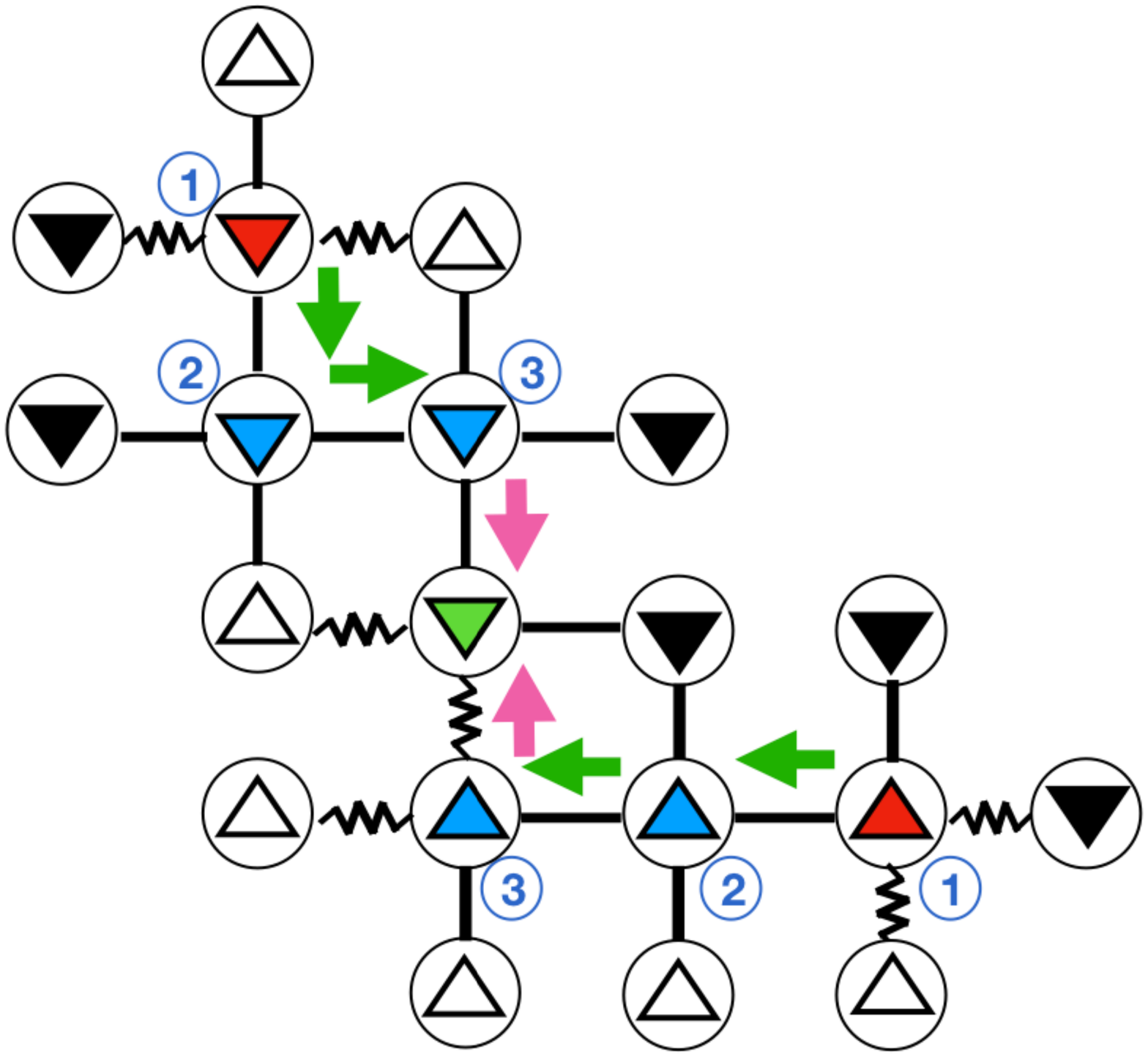}
\caption{ \label{pseudo3_sub2_2}
Visualization of type-3 dale minima with a split path in the reverse
direction of Fig.~\ref{pseudo3_sub2_1}.  The interactions $J_{ij} = -1$,
$1$, and $2$ are represented by wiggly lines (\protect\imgl{fig01}),
dashed lines (\protect\imgl{fig02}), and solid lines
(\protect\imgl{fig03}), respectively. White triangles
\protect\img{fig04} represent spin up and black triangles
\protect\img{fig05} represent spin down.  In the configuration
shown the two red triangles \protect\img{fig06} and
\protect\img{fig014} labeled \protect\img{fig07} are
zero-energy spins and can be flipped at no energy cost. Once they have
been flipped the spins drawn in blue \protect\img{fig09}
\protect\img{fig012} which are labeled
\protect\img{fig08} become zero-energy spins, which if flipped
allow the spins labeled \protect\img{fig010} to be flipped at no
energy cost. They form two paths, which are indicated by the green
arrows \protect\imga{fig011}. Only if both the paths have been
completed, i.e., both the spins labeled
\protect\img{fig010} have been flipped, the center spin indicated
in green \protect\img{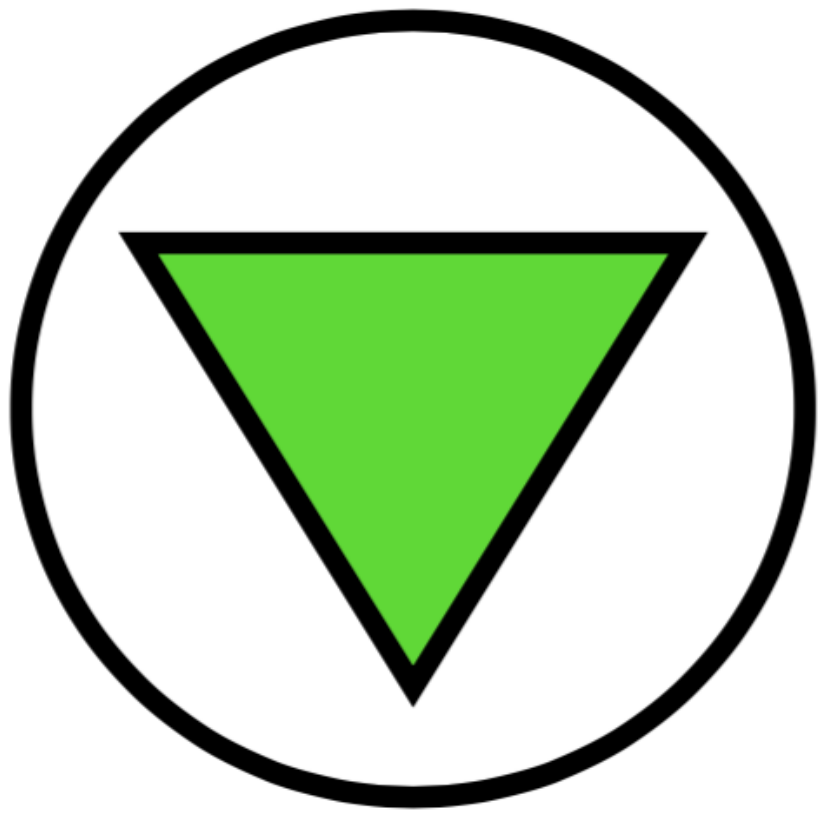} can be flipped at zero energy
cost. This double condition on the flippability of the center spin is
indicated by the pink arrows \protect\imga{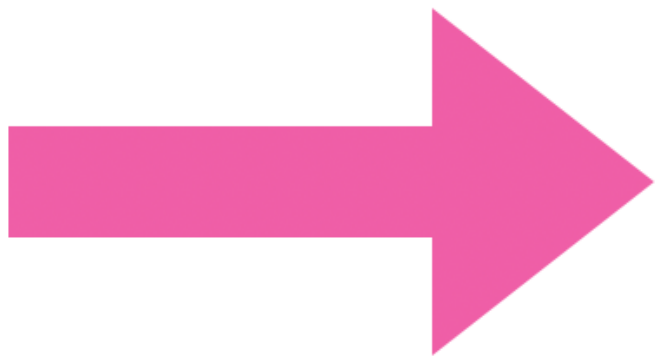}. }
\end{figure}

Figures~\ref{pseudo1} to \ref{pseudo3_sub2_2} illustrate via examples
the different dale minima. Upward triangles represent Ising spins with
the value $+1$ (up) and spins with the value $-1$ (down) are represented
by downward-pointing triangles. Zero-energy spins are drawn in red. The
interactions $J_{ij}$ between spins $i$ and $j$ are drawn as solid lines
when $J_{ij}=+2$, dashed when $J_{ij} = +1$, and wiggly lines when
$J_{ij} = -1$. Note that the examples given in Figs.~\ref{pseudo1} to
\ref{pseudo3_sub2_2}, only show the sections necessary to illustrate the
different types of dale minima and are part of larger systems.

Figure~\ref{pseudo1} shows an example of a type-1 dale minimum. In this
example, there are two zero-energy spins (red) As can be seen from the
figure, any combination of the two zero-energy spins (spin 1 and spin 2)
preserves the energy of the system. Without additional zero-energy spins
such a minimum has a degeneracy of $4$ originating from the free choice
of orientation, i.e., both spins up ($\uparrow\uparrow$), only one
of the spins up ($\uparrow\downarrow$ and $\downarrow\uparrow$) or both
spins down ($\downarrow\downarrow$). In this example the free choice of
flipping any combination of the zero-energy spins originates from their
separation, i.e., that they are not nearest neighbors. Therefore,
each choice of spin up or down does not affect the other, they are
independent of each other.

In Fig.~\ref{pseudo2} a type-2 dale minimum is illustrated. Here the two
zero-energy spins are nearest neighbors, which together with the
arrangement of the spin-spin interactions leads to the following
scenario: A flip of one spin influences the flipping of the other spin,
i.e., if spin 1 is flipped then spin 2 will have negative local energy
and thus can no longer be flipped without an increase in energy.
Similarly, if spin 2 is flipped, then spin 1 will in its up-configuration
be in a local minimum (has negative local energy) and can no longer be
flipped.  In this example the degeneracy originating from the
zero-energy spin effects is $3$, corresponding to the orientations of both
spins up ($\uparrow\uparrow$) or only one of them up and the other down
($\uparrow\downarrow$ and $\downarrow\uparrow$).  Note that in general,
a necessary condition for the occurrence of this type of dale minimum is
that at least two of the zero-energy spins have to be neighboring.

Figures~\ref{pseudo3_1} and \ref{pseudo3_2} illustrate a type-3 dale
minimum with a simple path.  In Fig.~\ref{pseudo3_1} only spin 1 is a
zero-energy spin. However, if spin 1 is flipped, then spin 2 becomes a
zero-energy spin and can be flipped without an increase in energy.
Further, if spin 1 and 2 are flipped, then spin 3 is a zero-energy spin
and can be flipped at zero energy cost leading to the configuration
depicted in Fig.~\ref{pseudo3_2}. Reversely, if the system starts in a
configuration of Fig.~\ref{pseudo3_2}, then the order of the flipping of
the spins is $3\rightarrow 2\rightarrow 1$, i.e., representing the
spins in the order $(1,2,3)$ the accessibility of the minima of this
dale can be depicted as
\begin{equation}
\uparrow\uparrow\uparrow ~ \longleftrightarrow ~
\downarrow\uparrow\uparrow ~ \longleftrightarrow
~\downarrow\downarrow\uparrow ~ \longleftrightarrow ~
\downarrow\downarrow\downarrow \>.
\label{schematic_pseudo3}\end{equation} 
Note that, while each of the configurations in the schematic
representation, Eq.~\eqref{schematic_pseudo3}, is a minimum on its own,
each can only access a limited number of neighboring minima belonging to
the same dale in the energy landscape. This means that the direct
accessibility of the minima is dependent on the current minimum. The
overall dale is pathway dependent. Pathway dependent in this example
means that there are only two pathways (one in each direction) that
connect all the minima belonging to the same dale minimum. In this
example, the local degeneracy of the dale minimum is $4$, representing
the four subminima.

Another variation of the type-3 dale minima is illustrated in
Figs.~\ref{pseudo3_sub2_1} and \ref{pseudo3_sub2_2}. We call this
subclass type-3 dale minimum with a split path, denoting that the
transition between the two extreme minima (i.e., minima in
Fig.~\ref{pseudo3_sub2_1} and minima in Fig.~\ref{pseudo3_sub2_2} in our
example) do not follow a single path but rather a path that can be
thought of to be split into two (or more) segments. A path here denotes
a sequence of flipping of spins which are neighboring to each other. In
the minima depicted in Fig.~\ref{pseudo3_sub2_1} only the center spin
(drawn in red) is a zero-energy spin and can be flipped at no cost.
However, once the center spin has been flipped it opens the two paths
denoted by 1, 2, and 3, which can be either flipped in sequence or
alternating between the two paths while maintaining the directional
order. A flip of the zero-energy spin makes spin labeled 1 a zero-energy
spin, which when flipped allows to flip spin number 2 and lastly spin
number 3, if 2 has been flipped. This directional order has to be
maintained in both directions in order for the transition to occur
without an increase in energy. Flipping the spins along both the paths
leads to the minimum depicted in Fig.~\ref{pseudo3_sub2_2}. The reverse
transition is shown in Fig.~\ref{pseudo3_sub2_2}. Here the two outermost
spins (depicted in red) are zero-energy spins and can be flipped at no
cost. This allows then the flipping of the spins marked with 2 and 3 in
subsequent order. Only after both of the paths have been completed,
i.e., if both the spins marked with 3 are flipped, the center spin
(drawn in green) becomes a zero-energy spin and can be flipped without
an increase in energy. The resulting minima after all the spin flips
have been completed is the minima depicted in Fig.~\ref{pseudo3_sub2_1}.
These transitions can also be visualized as
\begin{equation}
\begin{array}{ccccc}
~&~& \uparrow\uparrow\uparrow\uparrow\downarrow\downarrow\downarrow &~ &~\\
~&~& \updownarrow &~&~\\
~&~& \uparrow\uparrow\uparrow\downarrow\downarrow\downarrow\downarrow &~ &~\\
~& \nearrow\swarrow &~ & \searrow\nwarrow&~\\
\uparrow\uparrow\downarrow\downarrow\downarrow\downarrow\downarrow & ~&~&~& \uparrow\uparrow\uparrow\downarrow\uparrow\downarrow\downarrow \\
\updownarrow& ~ &~ & ~ & \updownarrow \\
\uparrow\downarrow\downarrow\downarrow\downarrow\downarrow\downarrow & ~&~&~& \uparrow\uparrow\uparrow\downarrow\uparrow\uparrow\downarrow \\
\updownarrow& ~ &~ & ~ & \updownarrow \\
\downarrow\downarrow\downarrow\downarrow\downarrow\downarrow\downarrow & ~&~&~& \uparrow\uparrow\uparrow\downarrow\uparrow\uparrow\uparrow \\
~& \searrow &~ & \swarrow&~\\
~&~& \curlyveedownarrow &~&~\\
~&~& \downarrow\downarrow\downarrow\downarrow\uparrow\uparrow\uparrow &~ &~
\end{array} \>,
\label{schematic_pseudo3sub2_1}\end{equation}
representing schematically the transition corresponding to
Fig.~\ref{pseudo3_sub2_1}, and
\begin{equation}
\begin{array}{ccccc}
~&~& \downarrow\downarrow\downarrow\downarrow\uparrow\uparrow\uparrow &~ &~\\
~& \nearrow\swarrow &~ & \searrow\nwarrow&~\\
\uparrow\downarrow\downarrow\downarrow\uparrow\uparrow\uparrow & ~&~&~& \downarrow\downarrow\downarrow\downarrow\uparrow\uparrow\downarrow \\
\updownarrow& ~ &~ & ~ & \updownarrow \\
\uparrow\uparrow\downarrow\downarrow\uparrow\uparrow\uparrow & ~&~&~& \downarrow\downarrow\downarrow\downarrow\uparrow\downarrow\downarrow \\
\updownarrow& ~ &~ & ~ & \updownarrow \\
\uparrow\uparrow\uparrow\downarrow\uparrow\uparrow\uparrow & ~&~&~& \downarrow\downarrow\downarrow\downarrow\downarrow\downarrow\downarrow \\
~& \searrow &~ & \swarrow&~\\
~&~& \curlyveedownarrow &~&~\\
~&~& \uparrow\uparrow\uparrow\downarrow\downarrow\downarrow\downarrow &~ &~\\
~&~& \updownarrow &~&~\\
~&~& \uparrow\uparrow\uparrow\uparrow\downarrow\downarrow\downarrow &~ &~
\end{array} \>,
\label{schematic_pseudo3sub2_2}
\end{equation}
representing schematically the transition depicted in
Fig.~\ref{pseudo3_sub2_2}. The wedged arrows denote the requirement that
both of the preceding paths have to be completed before the next step.
Note, that both the representations, Eq.~(\ref{schematic_pseudo3sub2_1})
and Eq.~(\ref{schematic_pseudo3sub2_2}), are simple representation of an
underlying highly-dimensional nature, meaning that the transitions
between the two outermost minima (the minima depicted in
Fig.~\ref{pseudo3_sub2_1} and Fig.~\ref{pseudo3_sub2_2}) have multiple
possible realizations corresponding to a large variety of pathways
across the dale in the energy landscape. Each step or point in the
configuration space is a subminimum situated within the dale. The dale
in our example is comprised of $17$ subminima, where the two minima
depicted in Fig.~\ref{pseudo3_sub2_1} and Fig.~\ref{pseudo3_sub2_2} are
its outermost points that can only be accessed from one minimum but have
two possible transitional minima for leaving in both the cases. The
paths shown here have been selected based on their significance in
highlighting the important aspects of the transitions.

\subsection{Disconnectivity Graphs}
\label{sec:discon}

In order to draw a disconnectivity graph, first, all the minima of the
system have to be calculated. We do this via a complete enumeration of
all possible spin configurations. Then, the height of the lowest energy
barrier connecting the minima is calculated by considering the possible
paths between the minima.  For this purpose the procedure of Cieplack
{\em et al.}~\cite{garstecki:99} for the calculation of possible
pathways was followed. Further, to deal with the large number of minima
in a computationally effective manner not all barriers
\cite{comment:barrier} are calculated but an approximation is used. This
approximation is based on the structural difference between the minima:
For each minimum only the barriers to the nearest $n$ minima are
calculated. With nearest, we mean the number of differences in the
orientation of the spins between any two minima and $n$ is a given
number of calculated barriers.  We reason that when searching for
low-energy pathways a system is likely to transition over intermediate
minima or funnels of intermediate minima. This approximation greatly
reduced the number of necessary barrier computations and thus allows for
a complete mapping of the energy landscape to a disconnectivity graph.
A similar approach was used in the analysis of potential energy
landscapes of hexapeptides~\cite{levy:98}. In general, the validity and
the extent to which this approximation can be applied depends on the
specific problem studied. We discuss this in more detail below in
Sec.~\ref{sec:model}.

\begin{figure*}
\includegraphics[width=0.7\linewidth, trim={0 0 0 0}, clip]{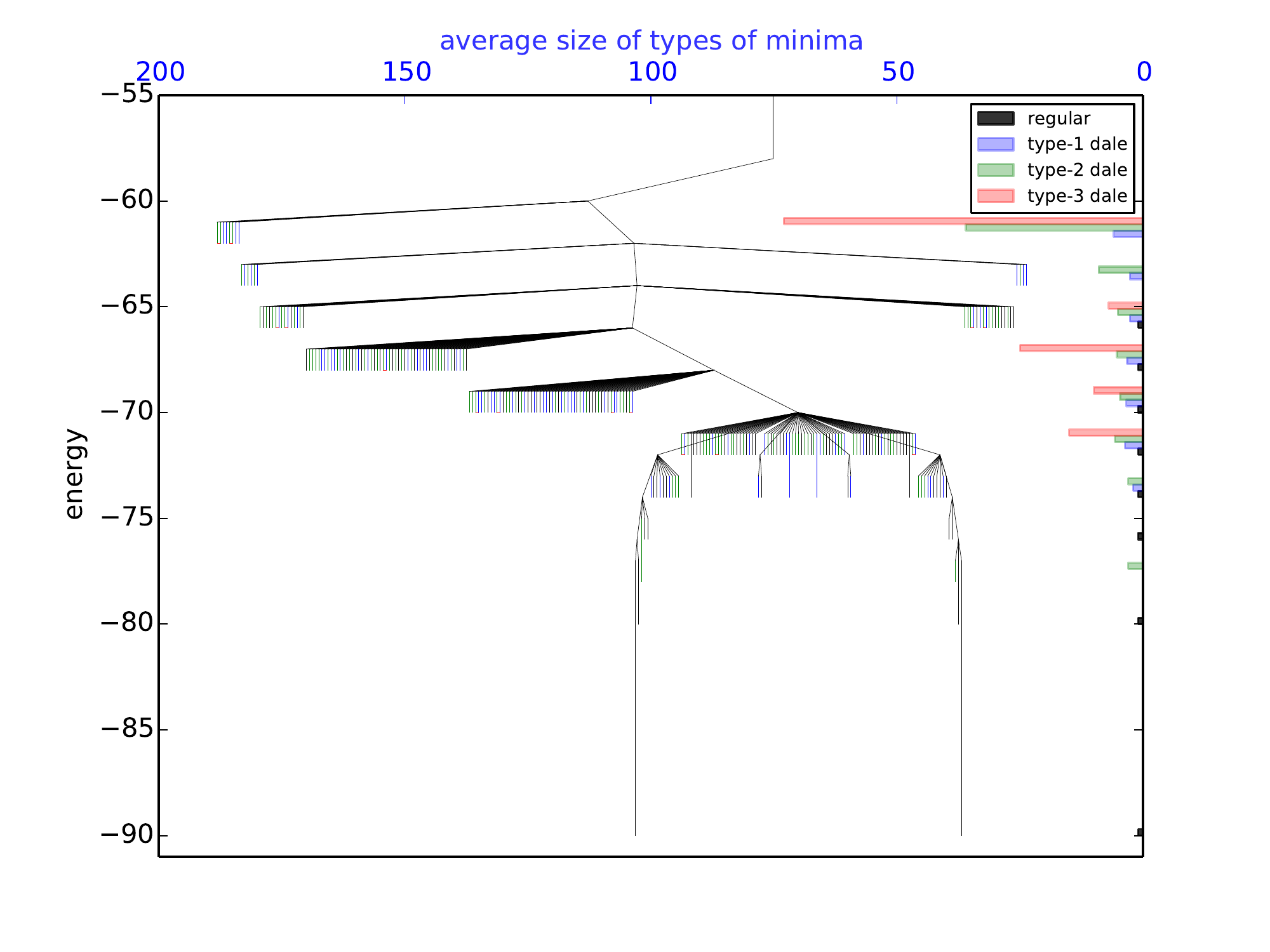}
\caption{\label{dis_graph_p1_1}
Example of a disconnectivity graph for a planted spin-glass model
($C_1$, see below) on a two-dimensional square lattice  with $N = 36$
spins.  The black vertical bars represent regular minima, the blue bars
represent type-1 dale minima, the green bars represent type-2 dale
minima and the red vertical bars indicate type-3 dale minima. The bar 
chart on the right shows the average size (i.e., the number of subminima 
comprising the dales) of the types of minima at their respective energy levels. 
The axis denoting the average size of types of minima only applies to 
the bar chart on the right-hand side of the disconnectivity graph. This axis does not 
represent the total number of minima in the system.}
\end{figure*}

\begin{figure*}
\includegraphics[width=0.7\linewidth, trim={0 0 0 0}, clip]{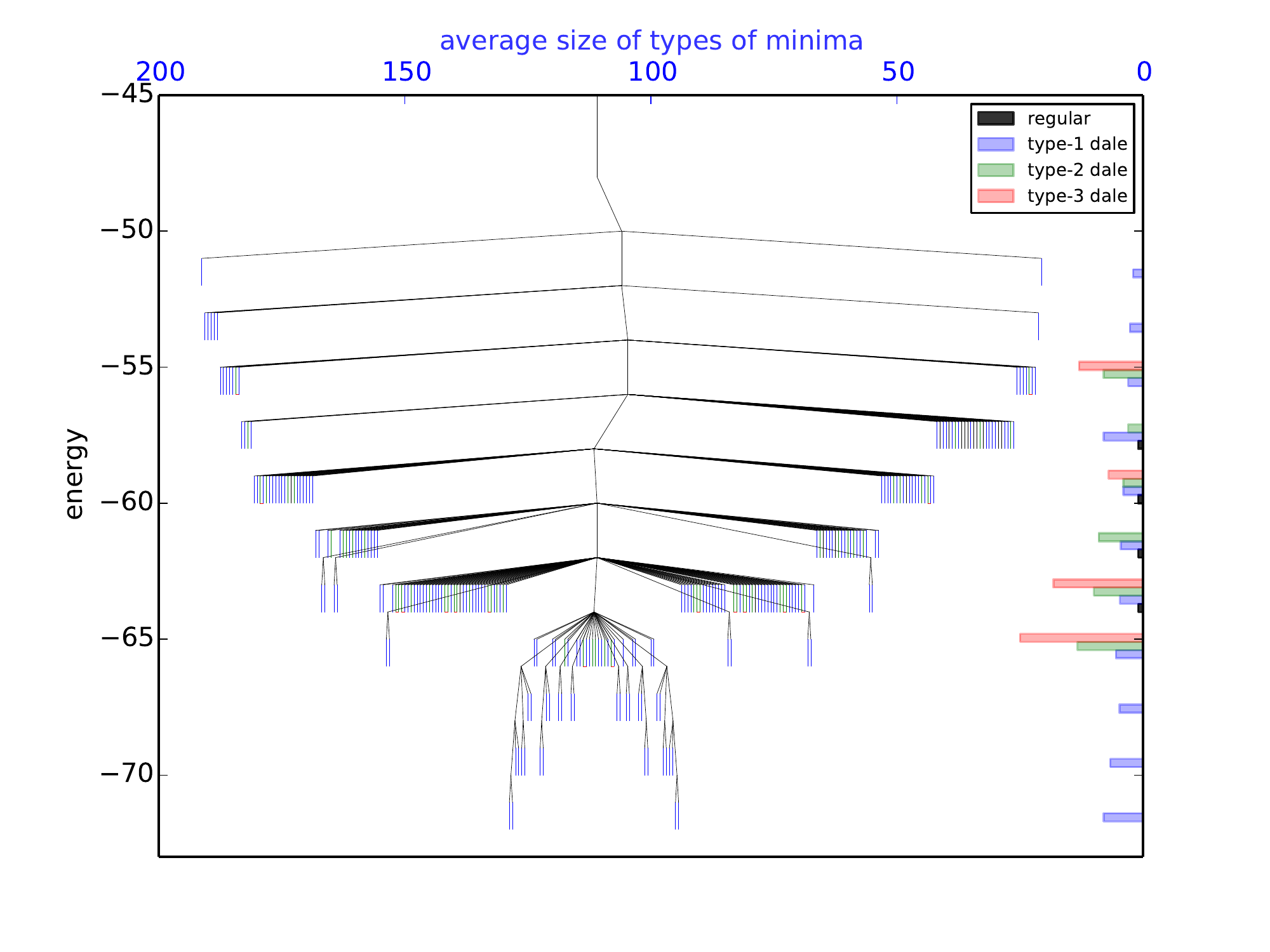}
\caption{\label{dis_graph_p2_1} 
Example of a disconnectivity graph for a planted spin-glass model
($C_2$, see below) on a two-dimensional square lattice  with $N = 36$
spins.  The black vertical bars represent regular minima, the blue bars
represent type-1 dale minima, the green bars represent type-2 dale
minima and the red vertical bars indicate type-3 dale minima. The bar chart on 
the right shows the average size of the types of minima at their respective 
energy levels. Note, that the axis denoting the average size of types of minima 
only applies to the bar chart on the right side of the disconnectivity graph.}
\end{figure*}

\begin{figure*}
\includegraphics[width=0.7\linewidth, trim={0 0 0 0}, clip]{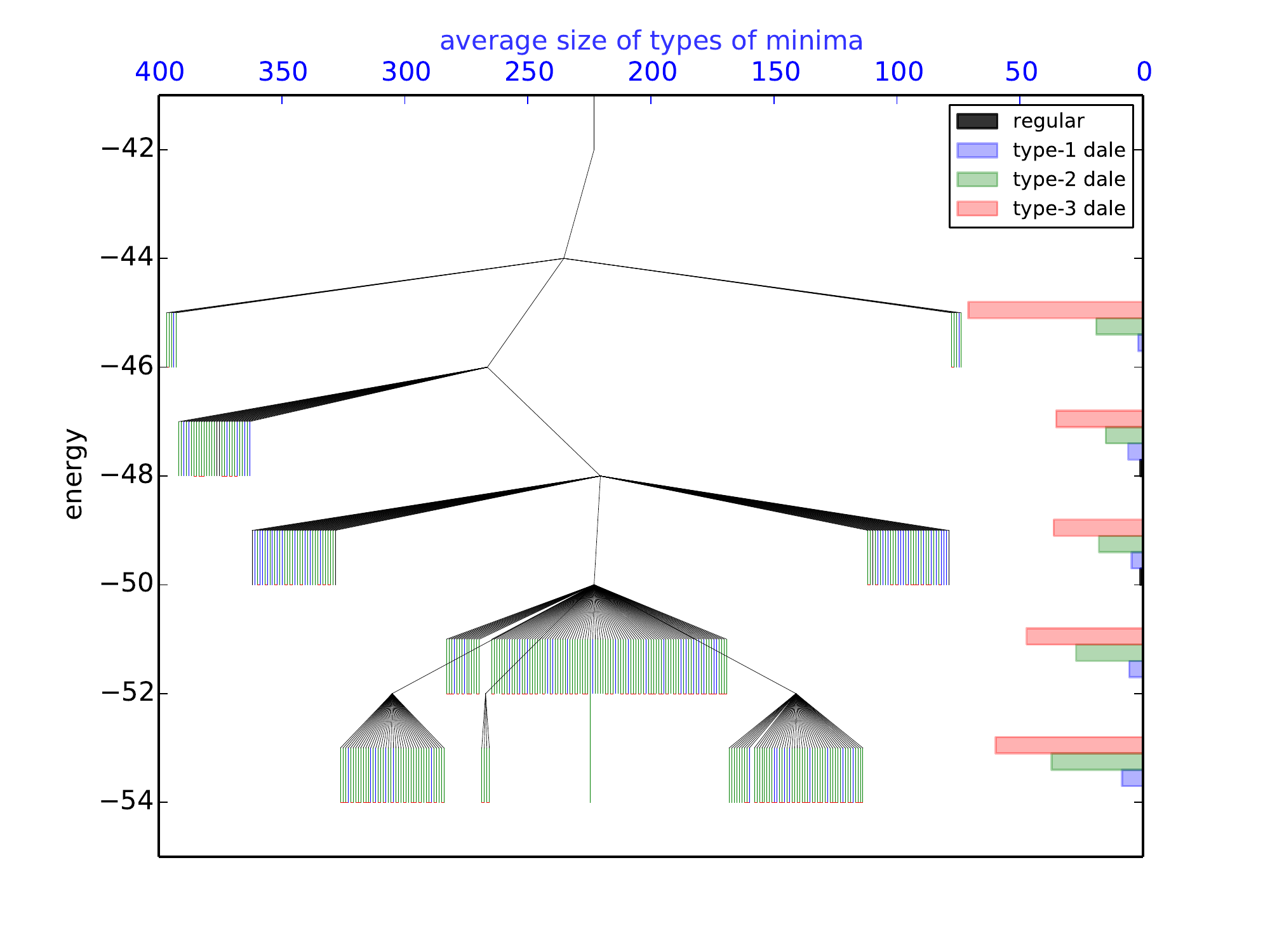}
\caption{\label{dis_graph_p3_1}
Example of a disconnectivity graph for a planted spin-glass model
($C_3$, see below) on a two-dimensional square lattice  with $N = 36$
spins.  The black vertical bars represent regular minima, the blue bars
represent type-1 dale minima, the green bars represent type-2 dale
minima and the red vertical bars indicate type-3 dale minima. The bar chart 
on the right shows the average size of the types of minima at their respective 
energy levels. Note, that the axis denoting the average size of types of minima 
only applies to the bar chart on the right side of the disconnectivity graph.}
\end{figure*}

Figures \ref{dis_graph_p1_1}, \ref{dis_graph_p2_1}, and
\ref{dis_graph_p3_1} show examples of disconnectivity graphs obtained
for the planted spin-glass models discussed below.  The minima are
represented by vertical bars, whose lowest points denote their energy.
The branching points represent the lowest high-energy barrier needed to
overcome to transition between the adjacent minima. The black bars
represent regular minima, i.e., minima for which a flip of any spin
would lead to an increase in energy.  The blue bars represent type-1
dale minima, the green bars represent type-2 dale minima, and red horizontal
lines joining two or more minima at the bottom indicate type-3 dale
minima.

The minima are joined and sorted into cluster structures according to
their energy barriers. Within each individual cluster, the minima are
arranged based on the number of spins up. Minima with the highest number
of spins up are drawn towards the left and sequentially minima with an
increasing number of spins down are arranged toward the
right~\cite{comment:structure}.

To indicate the size of the different types of minima, i.e., the number
of subminima belonging to the individual dales, we add a bar chart to
the right-hand side of the disconnectivity graph. This chart is arranged
such that the vertical axis represents the energies of the minima and
dales, and the horizontal axis represents the average number of
subminima belonging to the types, which we call the size of the types of
minima.

\section{Model}
\label{sec:model}

\begin{figure}[h!]
\includegraphics[width=0.8\linewidth,  trim={0 5cm 0 2cm}, clip]{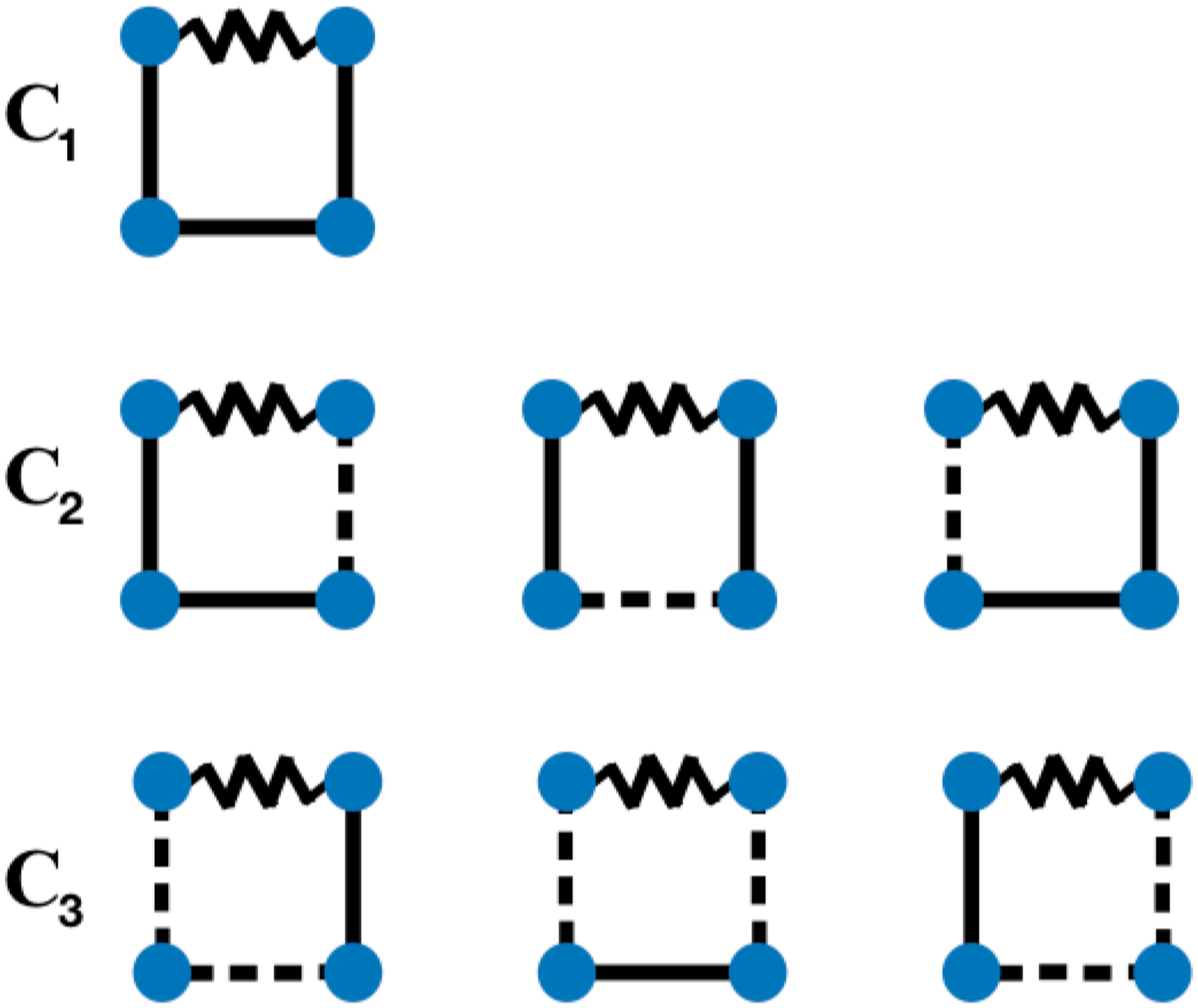}
\caption{\label{Cdraw1} 
Schematic representation of the planar unit cells $C_1$, $C_2$, and
$C_3$ in their invariant form. The solid lines (\protect\imgl{fig03})
represent $J_{ij}=+2$, the dashed lines (\protect\imgl{fig02})
represent $J_{ij}=+1$, and the wiggly lines (\protect\imgl{fig01})
$J_{ij} = -1$.}
\end{figure}

\begin{figure}[h!]
\includegraphics[width=0.8\linewidth,  trim={0 5cm 0 2cm}, clip]{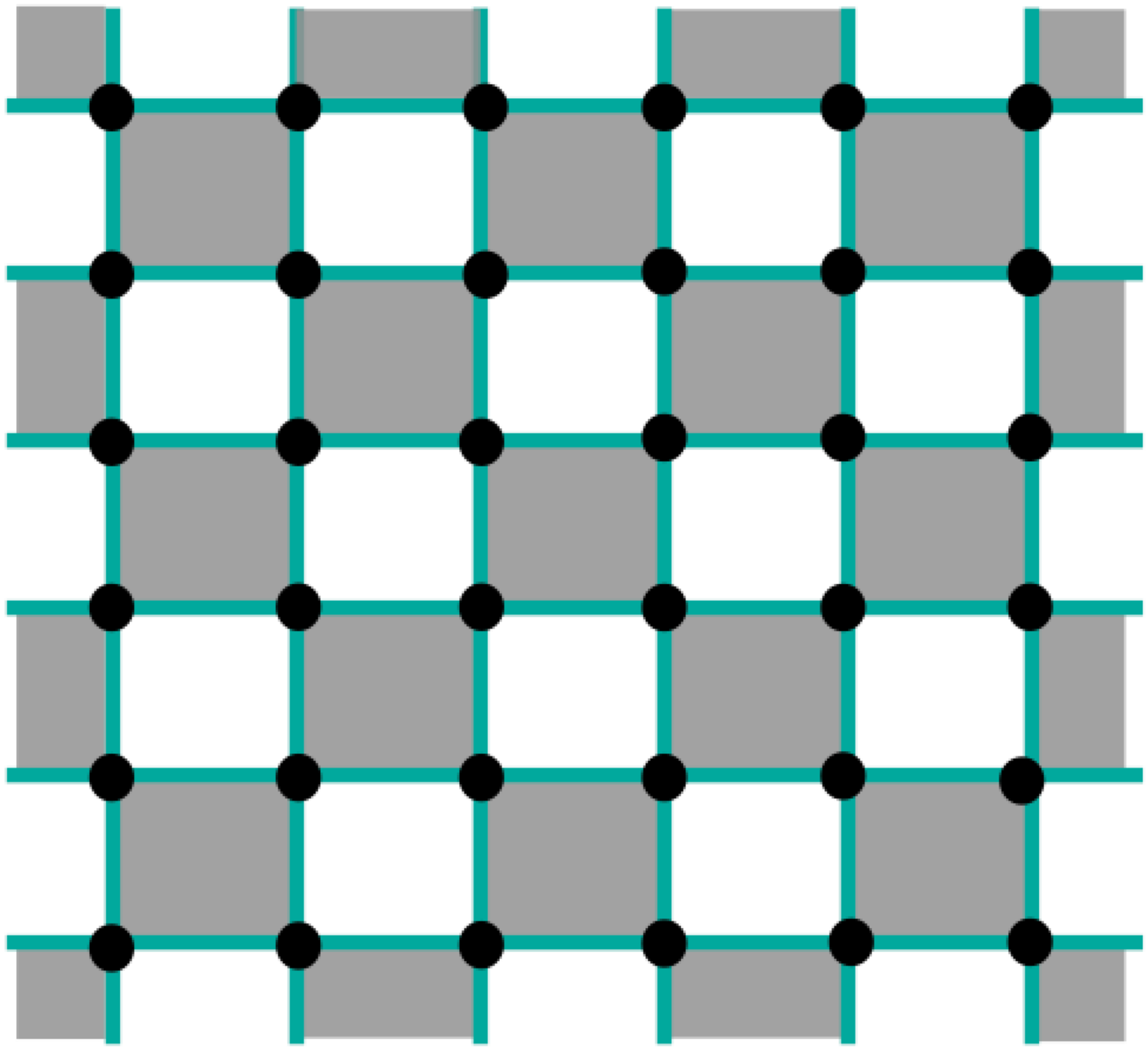}
\caption{\label{Cdraw2} 
Schematic representation of the construction pattern of a $6\times 6$
planar system with periodic boundaries. The filled squares represent the
tiles in which the unit cells are planted. }
\end{figure}

For the numerical experiments we focus on two-dimensional spin glasses
on a square lattice with tile-planted solutions consisting of
combinations of either $C_1$, $C_2$ or $C_3$ unit cells, as shown in
Fig.~\ref{Cdraw1} Unit cell $C_1$ consists of one $J_{ij}=-1$ and three
$J_{ij}=+2$ interactions between spins $i$ and $j$ that together form a
square and on whose vertices the spins $s_i$ are located. If all spins
are either oriented in the up or all in the down configuration the local
Hamiltonian of this configuration has ground-state energy
$H_{C_{1}}=-5$.  Unit cell $C_2$ consists of one $J_{ij}=-1$, one
$J_{ij}=+1$ and two $J_{ij}=+2$ spin-spin interactions, which can be
arranged in three distinct ways shown in Fig.~\ref{Cdraw1}. This unit
cell has four ground state configurations with an energy $H_{C_2}=-4$.
Unit cell $C_3$ has one $J_{ij}=-1$, two $J_{ij}=+1$ and one $J_{ij}=+2$
spin-spin interactions.  Their possible arrangements are shown in
Fig.~\ref{Cdraw1}. Unit cell $C_3$ has 6 ground states, corresponding to
an energy of $H_{C_{1}}=-3$. The total system is constructed by placing
the unit cells in a checker-board fashion such that only the vertices
are joined, see Fig.~\ref{Cdraw2}. Within a subgroup $C_1$, $C_2$, or
$C_3$ the unit cells are chosen and rotated randomly to form the larger
system.  Note that arranged in this way, the larger systems will have
two known ground-state solutions, where all spins are either pointing up
or all spins are pointing down. The total Hamiltonian is then given by
the sum over all interactions $J_{ij}$ centered around the individual spins
\begin{equation}
H=-\frac{1}{2}\sum_{\langle ij\rangle \in E}J_{ij} s_is_j 
= \frac{1}{2}\sum_{i} H_{E_i}s_i\>.
\label{hamil}\end{equation}
In Eq.~(\ref{hamil}) the sum is taken over all vertices,
$s_i=\pm1$ denote the values of the individual spins in their up or down
configurations, respectively, and $H_{E_i}$ is the local Hamiltonian of
the edges centered around spin $i$.

While some of the solutions of the ground states are constructed, each
of these types represents a model of different complexity. The hardness
in terms of time to solution (TTS) of pure and of mixed types of these
models has been studied numerically in Refs.~\cite{hamze:18,perera:19}.
In this work we map out the underlying energy landscapes of these
tile-planted systems.

\section{Results}
\label{sec:results}

Results have been obtained for systems with $N = 6\times 6$ spins, i.e.,
$18$ unit cells. To obtain sufficient statistics for each of the $C_1$,
$C_2$ and $C_3$ types of planted spin glasses a sample of $100$ systems
is randomly generated and evaluated. The minima of the systems are
obtained by complete enumeration and for each of the minima, barriers
are calculated to the closest $80$ minima. Although we only study one
system size in this work, the obtained results should be qualitatively
representative for the underlying behavior of the model system.

\begin{figure}[h!]
\includegraphics[width=\linewidth]{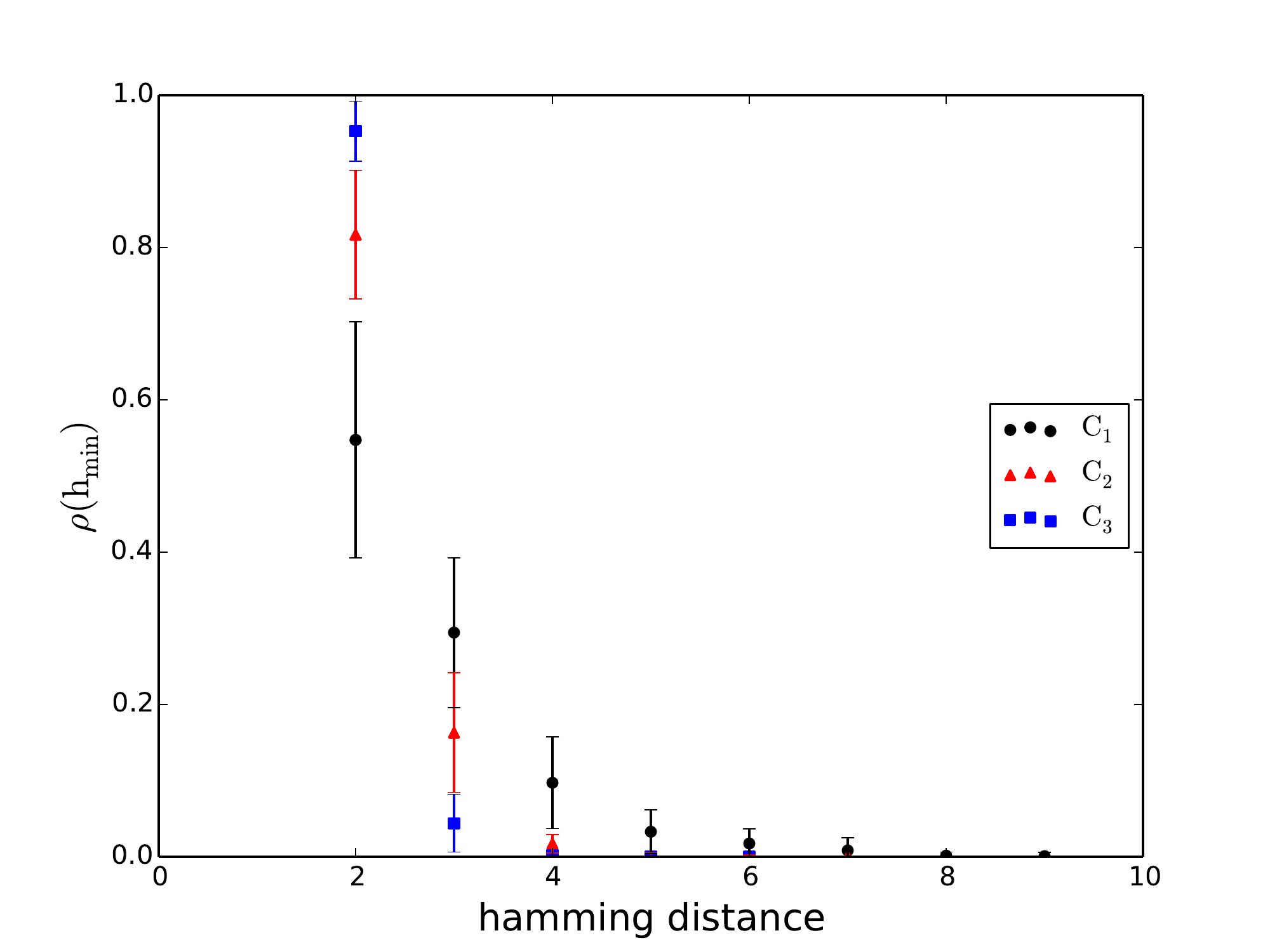}
\caption{\label{min_ham} 
Distribution of the minimum hamming distance for the $C_1$, $C_2$, and
$C_3$ models. The vertical bars denote the standard deviation.}
\end{figure}

Figure~\ref{min_ham} shows the distribution of the minimum hamming
distance for the three models of planted spin glasses. The hamming
distance here is understood as the minimum number of differences in the
orientation of spins between two minima. For type-1 and type-2 dale
minima, only the hamming distances between the two closest subminima
are represented, i.e., between the subminima of each dale with the
shortest hamming distance. To take into account the complexity in
transitioning between the subminima belonging to type-3 dale minima,
for this class only the hamming distances between the minima farthest
apart on the energy landscape have been taken into account. Since for
each of the models studied multiple samples were analyzed, each of which
varies in the total number of minima, the normalization is in two steps.
First, for each sample separately the distribution of the minimum
hamming distance is determined and normalized to unity.  Subsequently,
an average over samples is performed and the distributions normalized.
The standard deviation is computed for the average over samples as are statistically
more signifcant.

As can be seen from Fig.~\ref{min_ham}, for the three models studied,
the shortest hamming distance that occurred most often was $2$ with a
relative occurrence of $54.7\%$ for $C_1$-based models, $81.7\%$ for
$C_2$-based models, and $95.3\%$ for $C_3$-based models . Overall, the
$C_3$ model has a larger number of minima than $C_1$ and $C_2$, see
Tab.~\ref{table1}. This leads to an overall smaller difference in the
orientation of the spins between neighboring minima and therefore
shorter hamming distances. Note, that the configuration space consists
of all possible combinations of the orientations of the spins of the
systems and therefore has the same size for all the three models. This
requires the minima of the $C_1$ model --- which on average have fewer
minima than $C_2$ and $C_3$ --- to be more sparsely distributed between
the two outermost configurations consisting of either all spins up or
all spins down. The remaining minima are distributed between the extreme
cases of all spins up and all spins down in model $C_1$, thus explaining
why there are less states only two spin flips apart than for the other two models,
and the rest requiring three ($29.4\%$), four ($9.7\%$), up to a maximum
of nine required spin flips between neighboring minima ($0.085\%$).
While these numbers will change with different system sizes, we estimate that 
the generic trend of the data where two spin flips are dominant remains.  The
maximum number of required spin flips between neighboring minima for the
$C_2$ model was found to be seven in our sample of 100 systems with an
occurrence of $0.018\%$, and for $C_3$ it is six at $0.007\%$.

\begin{figure}
\includegraphics[width=\linewidth,  trim={1cm 0 2cm 0}, clip]{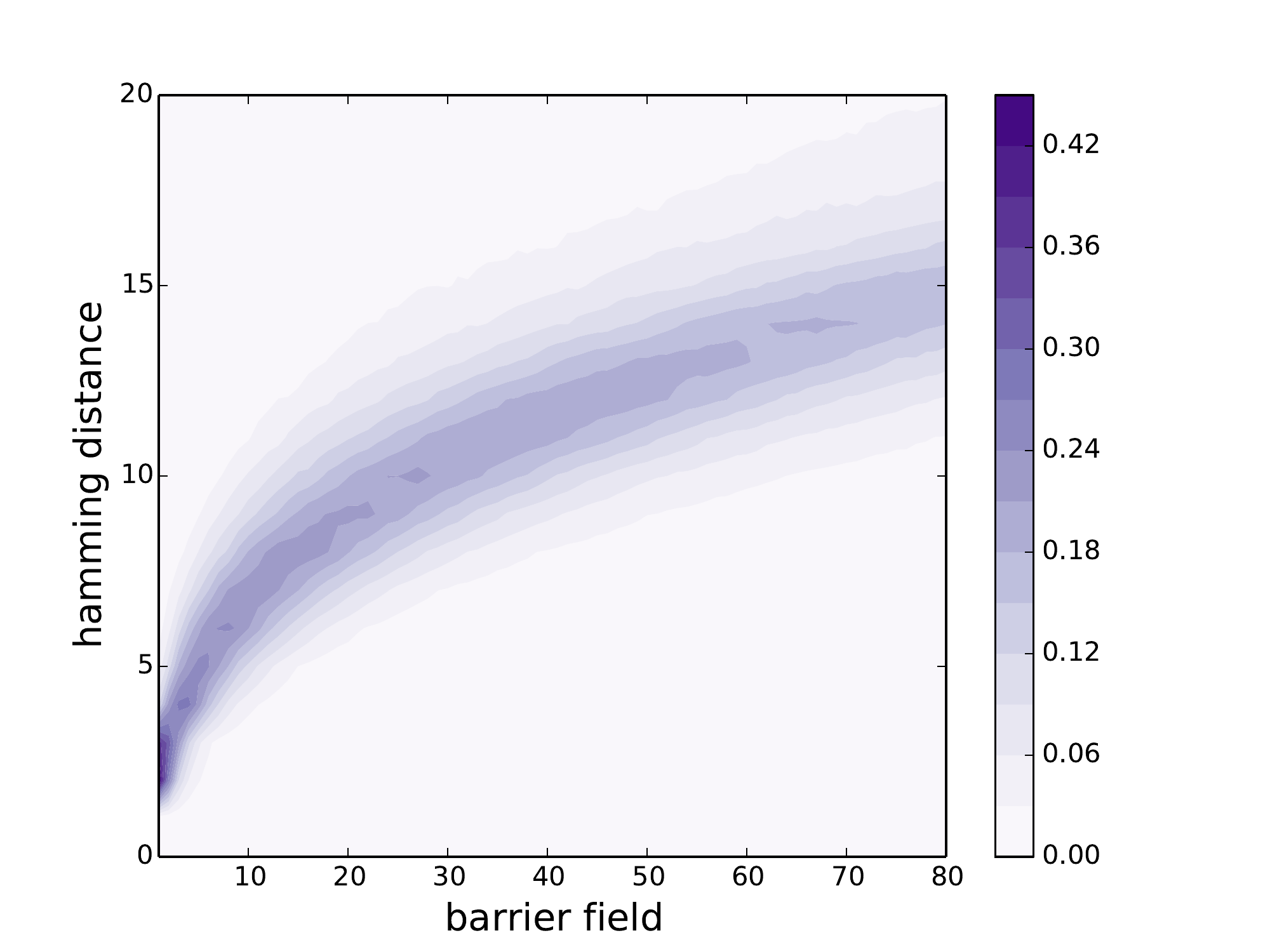}
\caption{\label{ham_bar_p1_1} Distribution of hamming distances vs the 
barrier field per minima of the $C_1$ model.}
\end{figure}

\begin{figure}
\includegraphics[width=\linewidth,  trim={1cm 0 2cm 0}, clip]{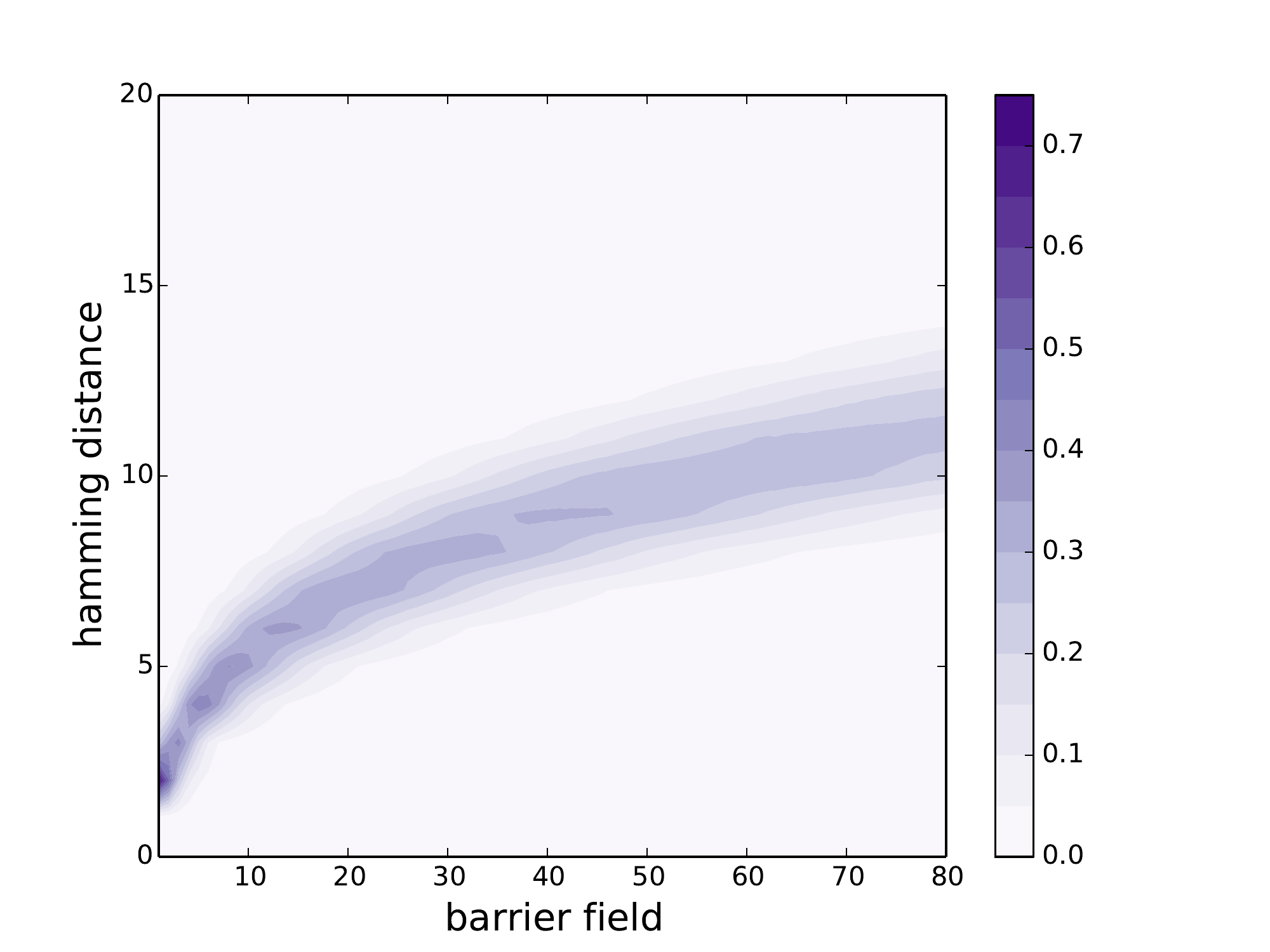}
\caption{\label{ham_bar_p2_1} Distribution of hamming distances vs the 
barrier field per minima of the $C_2$ model.}
\end{figure}

\begin{figure}
\includegraphics[width=\linewidth,  trim={1cm 0 2cm 0}, clip]{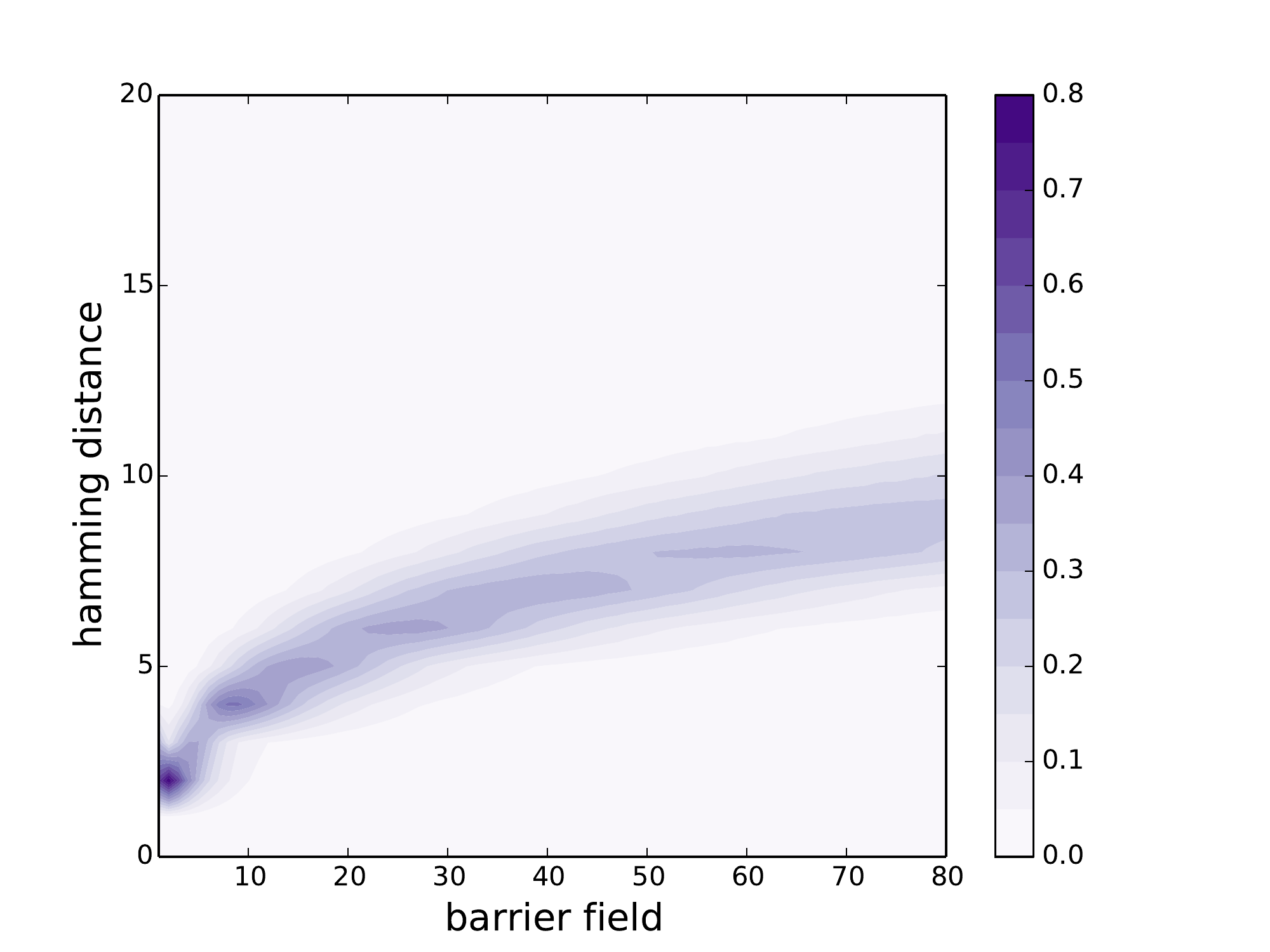}
\caption{\label{ham_bar_p3_1} Distribution of hamming distances vs the 
barrier field per minima of the $C_3$ model.}
\end{figure}

To verify the validity of our approximation of including only up to $80$
barriers for each minimum, we introduce a {\em barrier field.} The
barrier field is the storage allocation of the barriers which are sorted
according to their hamming distances, i.e., the hamming distances are
sorted according to their size and allocated to the hamming field such
that small hamming field values store the information about the barriers
for the smallest hamming distances.  Figures~\ref{ham_bar_p1_1},
\ref{ham_bar_p2_1}, and \ref{ham_bar_p3_1} show the distribution of the
hamming distances for the barrier field for the $C_1$, $C_2$ and the
$C_3$ model systems, respectively. The figures show the results
normalized for each value of the barrier field separately and give an
indication of the sizes of the hamming distances that are covered by
calculating barriers including only $80$ neighboring states.

As can be seen from Figs.~~\ref{ham_bar_p1_1}, \ref{ham_bar_p2_1}, and
\ref{ham_bar_p3_1} , in all of the model systems, only a few allocations
(i.e., $<10$) are necessary to include all the minima within the
shortest hamming distance. By obtaining just about $50$ barriers for
each of the minima hamming distances up to $26$ spin flips are covered
for the $C_1$ model system with the highest occurrence of $12$ or $13$
required flips within the $50$th's barrier calculation. For the $C_2$
model system hamming distances up to $15$ occur, with $9$ or $10$ spin
flips having the highest occurrence. For the $C_3$ model system hamming
distances of up to $16$ occur, with the highest occurrence of $7$ and
$8$ necessary flips in the $50$ barrier calculation.  The difference in
the covered hamming distances of the models is due to the difference in
the number of minima between the different models, see
Tab.~\ref{table1}.

\begin{figure}
\includegraphics[width=\linewidth, trim={1cm 0 2cm 0}, clip]{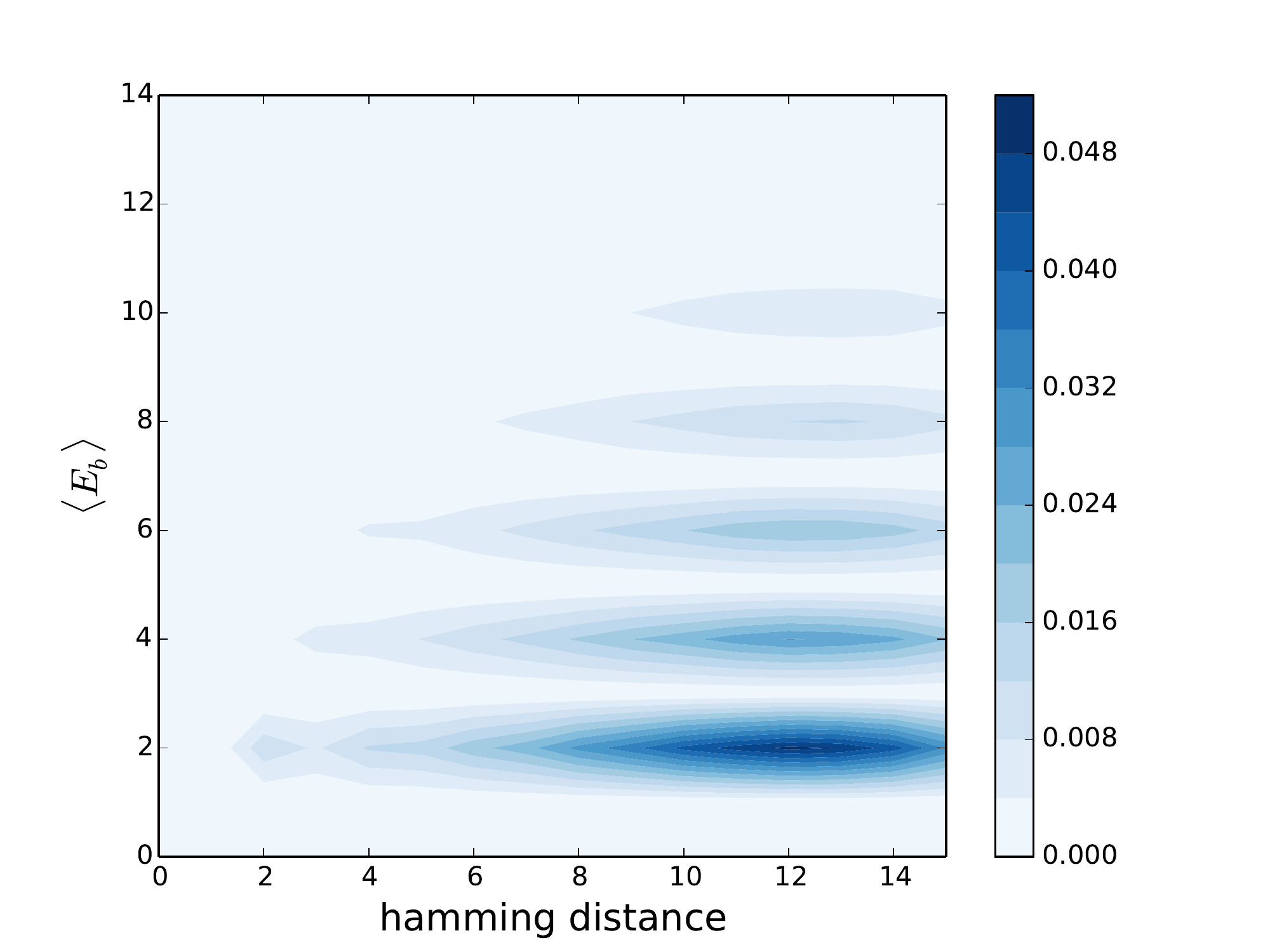}
\caption{\label{bar_hamdis_p1_1} Distribution of the barrier heights versus the hamming distance for the $C_1$ model. }
\end{figure}

\begin{figure}
\includegraphics[width=\linewidth,  trim={1cm 0 2cm 0}, clip]{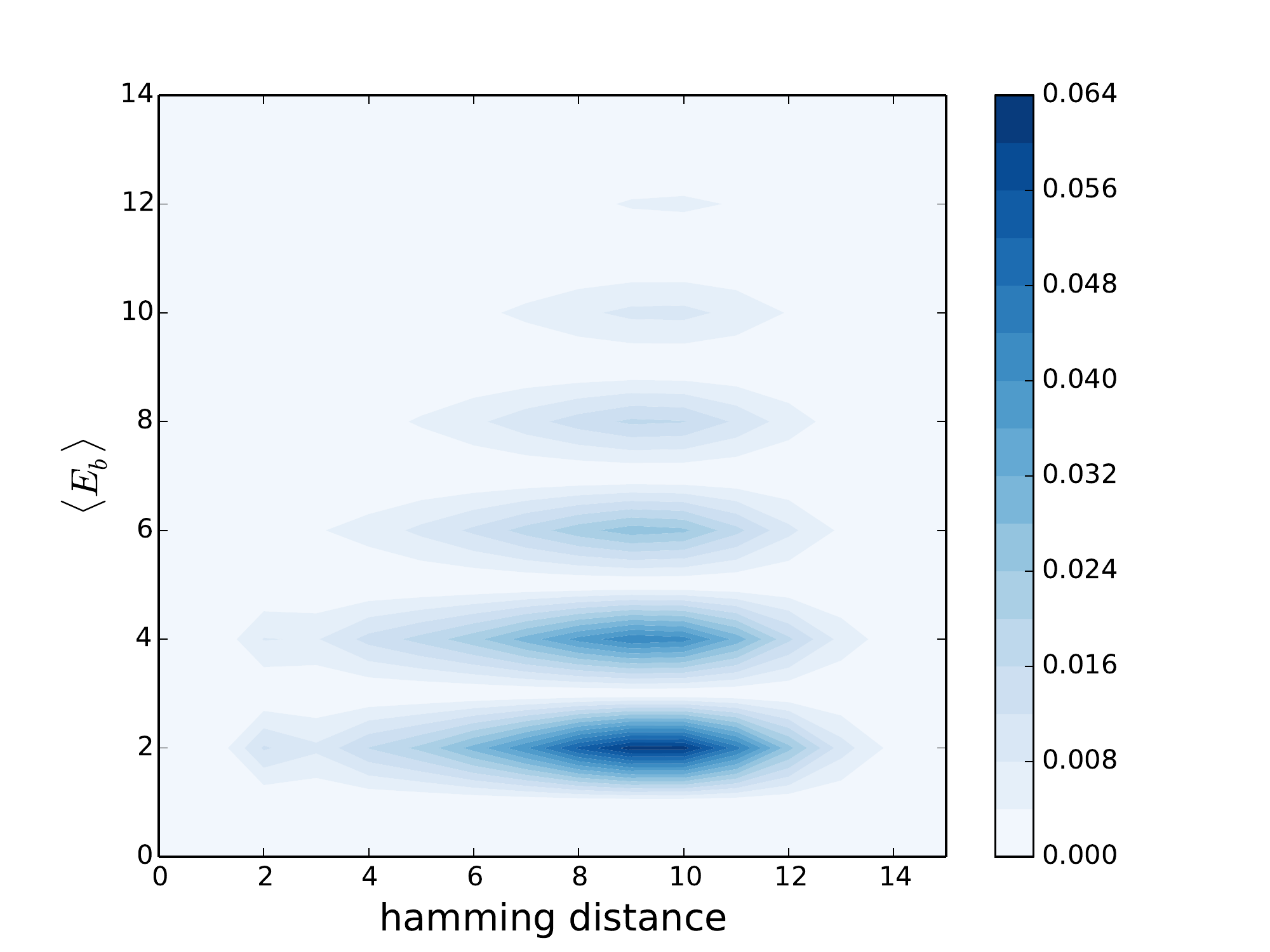}
\caption{\label{bar_hamdis_p2_1} Distribution of the barrier heights versus the hamming distance for the $C_2$ model. }
\end{figure}

\begin{figure}
\includegraphics[width=\linewidth,  trim={1cm 0 2cm 0}, clip]{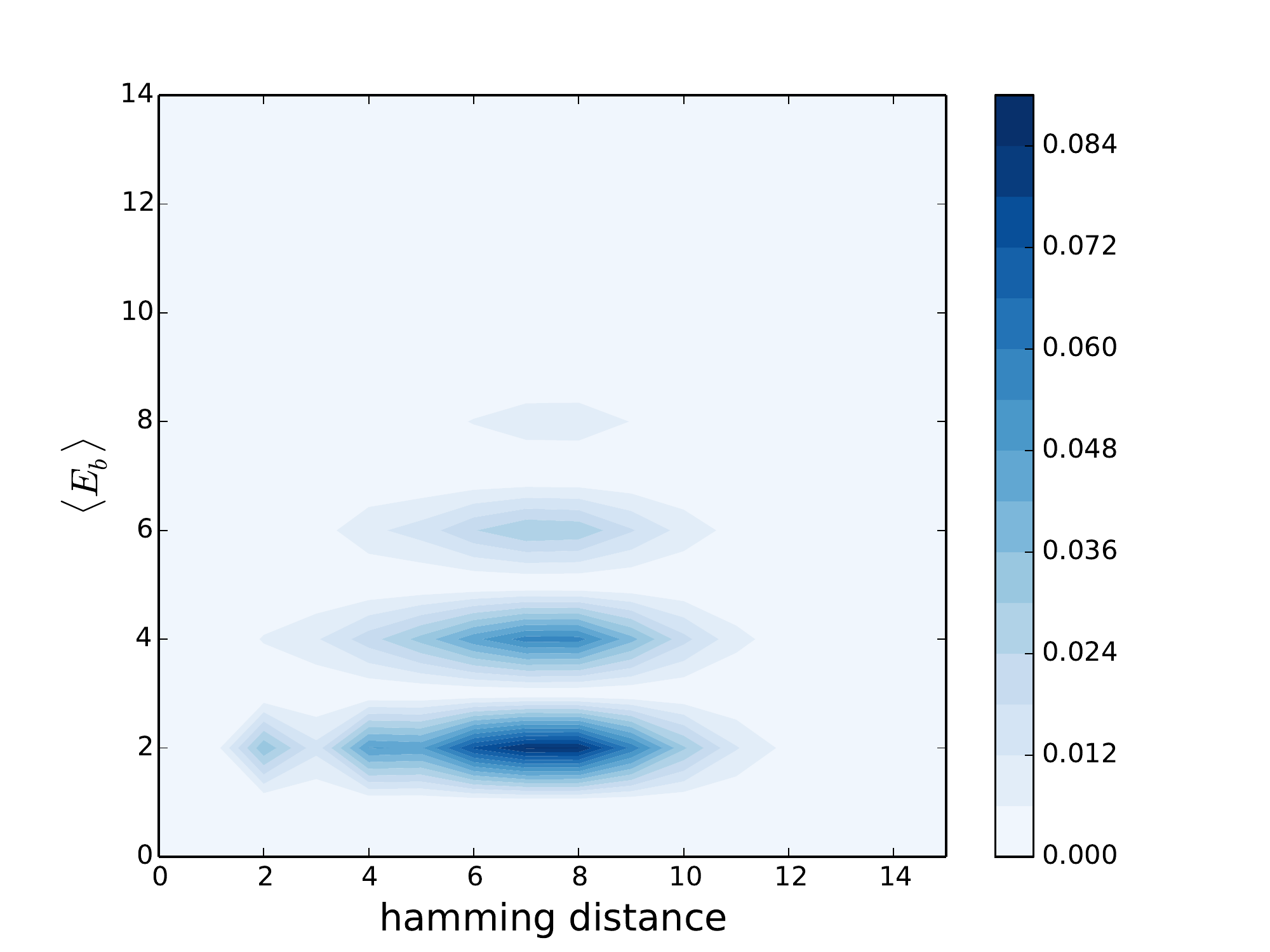}
\caption{\label{bar_hamdis_p3_1} Distribution of the barrier heights versus the hamming distance for the $C_3$ model. }
\end{figure}

Figures~\ref{bar_hamdis_p1_1}, \ref{bar_hamdis_p2_1}, and
\ref{bar_hamdis_p3_1} show the distribution of the average barrier
heights versus the hamming distance of the three model systems $C_1$,
$C_2$, and $C_3$. They visualize the calculated range of the barrier
field.  Note that a dale minimum consists of multiple minima that are
joined by zero-energy spin flips, i.e., they form dales on the energy
landscape on which the system can transit without an increase in energy.
In order to avoid unnecessary double counting, only the barriers
corresponding to the shortest hamming distance of minima, belonging to
the same dales for type-1 dales and type-2 dales, have been taken into
account.  Due to the dependence on the pathway, for type-3 dale minima
the hamming distance is determined between the endpoints of the
individual dales. The figures are normalized to unity for the full
distribution.

As can be seen, in all model systems, the barrier heights for small
hamming distances are low. This is due to the fact that when only a few
spin flips are necessary to change between minimum energy configurations
then only a few spin-spin interactions contribute to the increase in
energy of the system. Minima separated by larger hamming distances show
a larger distribution of the energy barriers but are also still
dominated by small barrier heights. Furthermore, comparing
Figs.~\ref{bar_hamdis_p1_1}, \ref{bar_hamdis_p2_1}, and
\ref{bar_hamdis_p3_1}, a significant contribution of high-energy
barriers occurs only at intermediate hamming distances for the model
systems $C_2$ and $C_3$. The energy of the highest barrier is larger for
$C_2$ than for $C_1$ and $C_3$, with $C_3$ having the lowest high
barrier.  The results from $C_3$ originate from two factors. First, due
to the large number and huge degeneracy of dale minima in the $C_3$
model system a larger section of the configuration space is covered than
for the other models. Thus, on average a smaller number of spin flips is
necessary to traverse between the minima considered in the barrier
calculation, also explaining the shorter hamming distances.  Second, the
$C_3$ model system has more spin-spin interactions of smaller value than
the two other models, which leads to that even if a barrier has to be
crossed, the necessary increase in energy between two minima is, in
general, smaller than for the other two models. This explains the
difference in the observed barrier heights.

These together with the distribution of the hamming distance for the
barrier field (see Figs. ~\ref{ham_bar_p1_1},
\ref{ham_bar_p2_1} and \ref{ham_bar_p3_1}) the distribution of the
energy barriers confirm with our approximation of the calculation of the
barriers.  This approximation is based on the assumption that minimum
energy transitions between minima, which are far apart on the energy
landscape, will most likely occur through pathways that are leading
along or through dales of intermediate minima. During such a transition
the necessary increase in energy for the total transition will be
already captured by intermediate barriers.  While low-energy barriers
had the highest occurrence indicating the high arrangement of minima
into first-order cluster structures, our reduced computation of barriers
(i.e., up to a maximum of $80$ barriers for each minimum) also
covered higher-energy transitions. This allows for transitions out of
the local cluster structures. It indicates that for each local cluster
there are minima for which the calculated barriers include high-energy
transitions, and hence enables the overall connectivity of the minima.

\begin{table}
\caption{\label{table1}
Number of the minima types $\mathcal{N}$ in the reduced description, percentage of the minima $\% \mathcal{N}$ in the reduced description, average size of minima types $\mathcal{S}$, actual number of minima types $\mathcal{N}^{*}$ , percentage of the actual number of minima $\% \mathcal{N}^{*}$. The values are averages obtained from our samples of 100 systems for each of the $C_1$, $C_2$ and $C_3$ models of tile-planted spin glasses. R denotes regular minima, D1, D2, and D3 denote type-1, type-2, and type-3 dale minima, respectively 
\cite{comment:table}.}
\begin{tabular*}{\columnwidth}{@{\extracolsep{\fill}} l r r r r}
\hline
\hline
~~& regular & type-1 & type-2 & type-3 \\
\hline
\hline
$\bf C_1$ & ~ & ~ & ~ & ~\\
\hline
$\mathcal{N}$ & $70.7 $ & $67.4$ & $69.3$ & $15.3$ \\
$\% \mathcal{N}$ & 34.1 & 32.5 & 33.4 & ~ \\
$\mathcal{S}$ & 1.0 & 2.8 & 8.4 & 16.8\\
$\mathcal{N}^{*}$ & $70.7 $ & $188.7$ & $582.1$ & $257.0$ \\
$\% \mathcal{N}^{*}$ & 8.4 & 22.4 & 69.2 & ~\\ 
\hline
$\bf C_2$ & ~ & ~ & ~ & ~\\
\hline
$\mathcal{N}$ & $147.6 $ & $238.7$ & $127.9$ & $28.1$ \\
$\% \mathcal{N}$ & 28.7 & 46.4 & 24.9 & ~\\
$\mathcal{S}$ & 1.0 & 3.1 & 7.1 & 12.7\\
$\mathcal{N}^{*}$ & $147.6 $ & $740.0$ & $908.1$ & $356.9$ \\
$\% \mathcal{N}^{*}$ & 8.2 & 41.2 & 50.6 & ~\\ 
\hline
$\bf C_3$ & ~ & ~ & ~ & ~\\
\hline
$\mathcal{N}$ & $61.1 $ & $298.6$ & $415.2$ & $102.1$ \\
$\% \mathcal{N}$ & 7.9 & 38.5 & 53.6 & ~\\
$\mathcal{S}$ & 1.0 & 4.8 & 17.1 & 30.1\\
$\mathcal{N}^{*}$ & $61.1 $ & $1433.3$ & $7099.9$ & $3073.2$ \\
$\% \mathcal{N}^{*}$ & 0.7 & 16.7 & 82.6 & ~\\ 
\hline 
\hline 
\end{tabular*}
\end{table}

Table~\ref{table1} shows the number of minima for the different types of
minima of the $C_1$, $C_2$ and the $C_3$ model system. As can be seen
from the table, the actual number of minima $\mathcal{N}^{*}$ is much
higher than the number of minima in the reduced description
$\mathcal{N}$. This is especially striking for the $C_3$ model, whose
energy landscape exhibits a huge number of type-2 dale minima with large
sizes of energy dales.  The size $\mathcal{S}$ gives the average number
of minima belonging to a dale. The average total number of minima,
counting each state on a minimum energy dale as individual minimum, is
$856.8$ for $C_1$, $1823.8$ for $C_2$ and $8696.4$ for $C_3$ (post
sample average). Especially for $C_3$, drawing these huge numbers of
minima into disconnectivity graphs would lead to problems. However,
using our classification scheme, the average total number of minima in
the reduced description is $207.4$ for $C_1$, $514.2$ for $C_2$ and
$774.9$ for $C_3$, thus making it much simpler to display the energy
structure of the minima in the disconnectivity graphs. In the reduced
description of the minima, in the $C_1$ model regular minima and type-1
and type-2 dale minima seem to have equally likely occurred.  However,
this observation is to be taken with precaution since the reduced
description does not take the size of the dales into consideration.
Taking the size $\mathcal{S}$ of the minimum energy dales into account,
one can see from Tab.~\ref{table1}, that most of the individual minimum
energy configurations belong to the type-2 dale minima.  Similarly, for
the $C_2$ model, while in the reduced description the number of type-1
dale minima dominates, taking into account the different sizes
$\mathcal{S}$ of the dales shows that the majority of the individual
minimum energy configurations belong to the type-2 dale category.

Lastly, compared to the other two model systems, the $C_3$ model has the
smallest percentage of regular minima and the highest percentage of
type-2 dale minima. Together with the high occurrence of low-energy
states, this indicates that the energy landscape of this model is easier
to traverse using standard optimization routines than the other two
models.

\begin{figure}
\includegraphics[width=\linewidth]{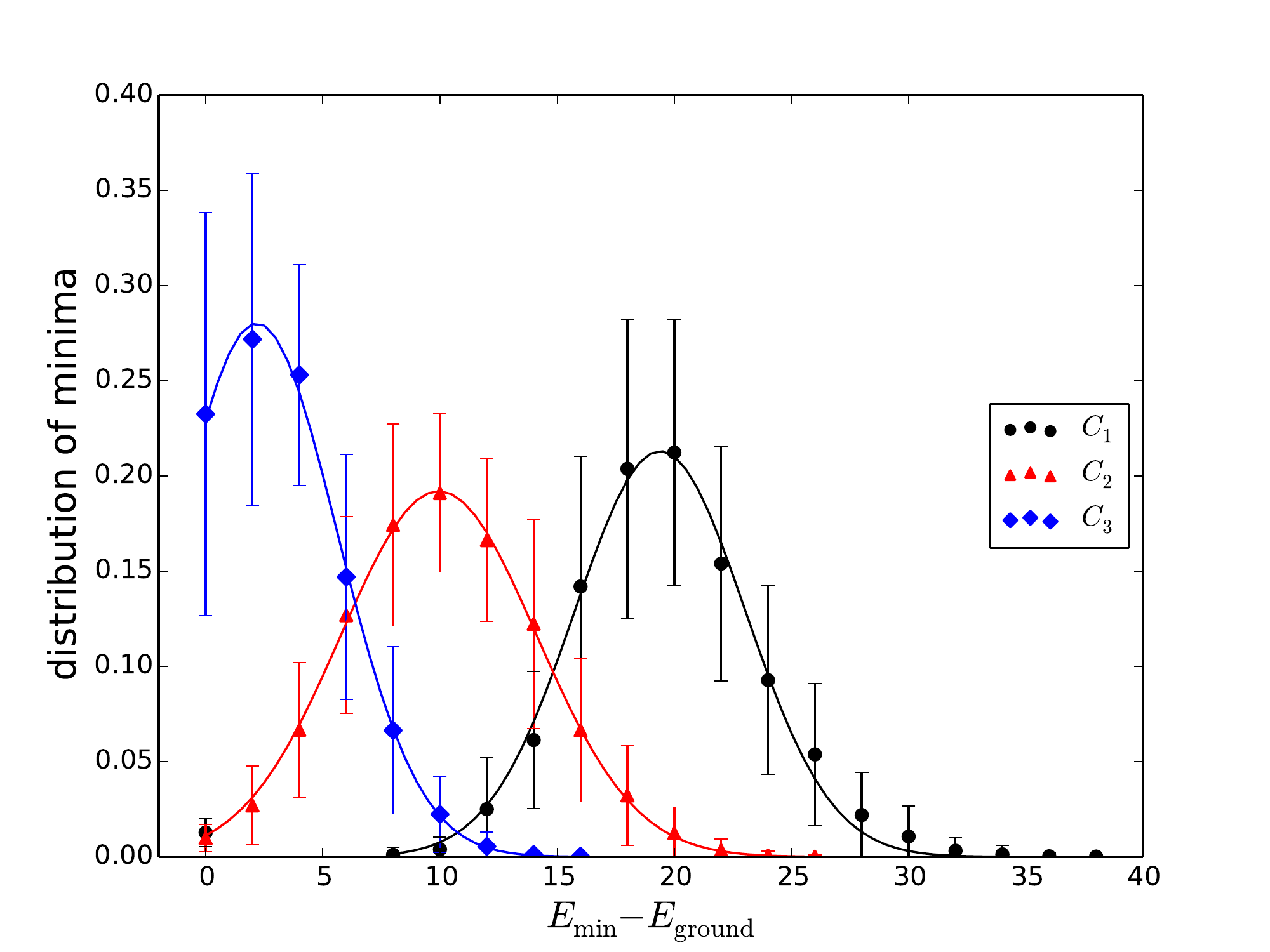}
\caption{\label{dis_min}
Distribution of the energy of the minima relative to the energy of the
ground state for the different planted spin-glass model systems $C_1$
(black circles), $C_2$ (red triangles) and $C_3$ (blue diamonds). The
lines are Gaussian fits and are guides to the eye}
\end{figure}

Figure~\ref{dis_min} shows the average distribution of the energy of the
minima $E_{\mathrm{min}}$ in the reduced description and their standard
deviation relative to the ground state energy $E_{\mathrm{ground}}$ for
the three models of tile-planted spin glasses.  The large standard
deviation  in the data is due to the large variation in the number of
minima in the generated systems of the three models.  As can be seen
from Fig.~\ref{dis_min} and comparing with the disconnectivity graphs in
Figs.~\ref{dis_graph_p1_1}, \ref{dis_graph_p2_1}, and
\ref{dis_graph_p3_1}, the $C_1$ model system is distinguished by a large
gap between the ground state and higher-order minima. This is not the
case for the two other model systems.  The majority of the minima of the
$C_2$ model system are found at medium energies, whereas the $C_3$ model
is distinguished by a large number of minima energetically close to the
ground state.

\begin{figure}
\includegraphics[width=0.6\linewidth]{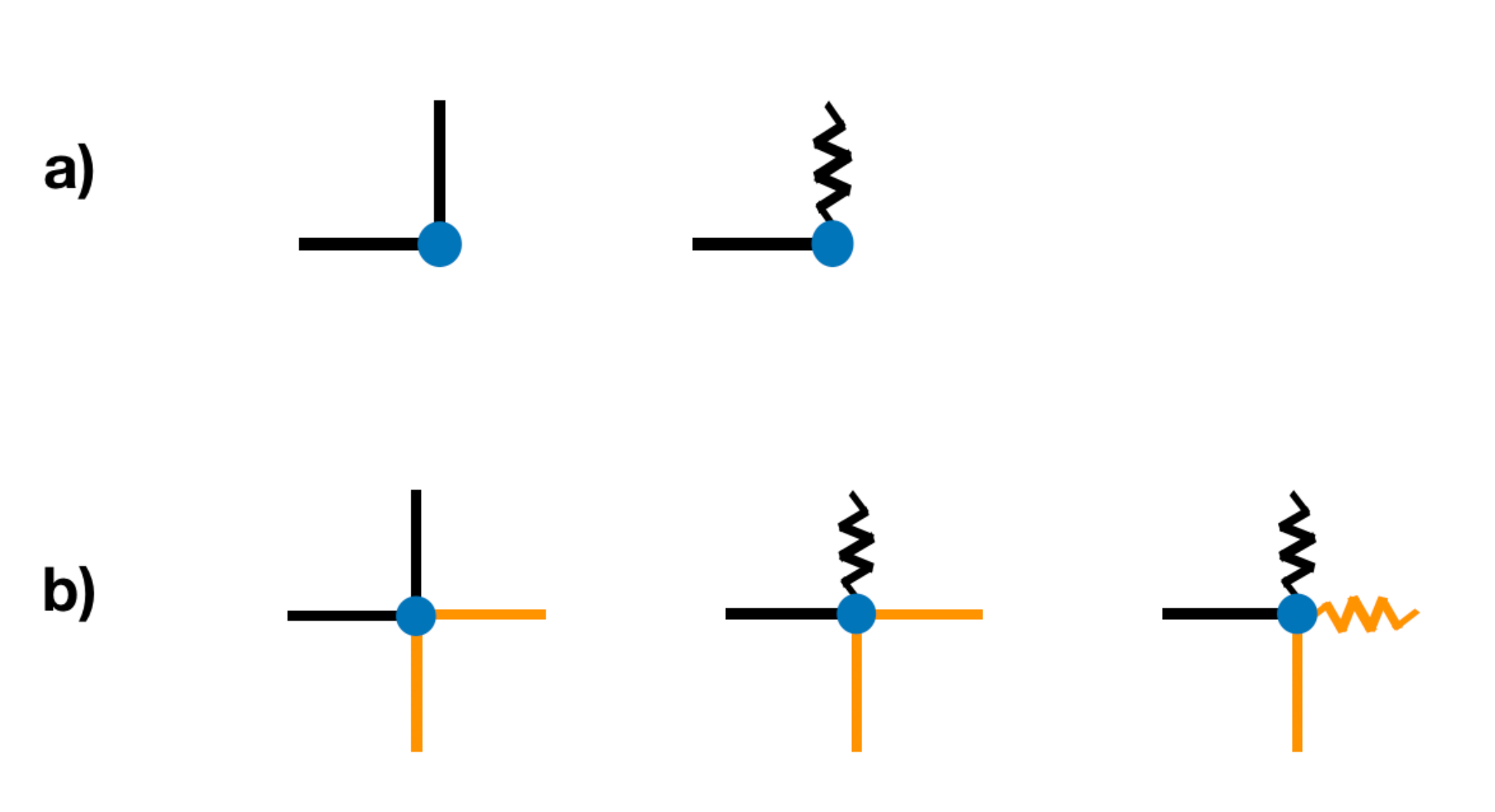}
\caption{\label{C1_planted} 
(a) Possible contributions of spin-spin interactions of an individual
$C_1$ unit cells to a vertex. The solid lines (\protect\imgl{fig03})
represent $J_{ij}=+2$ spin-spin interactions and the wiggly lines
(\protect\imgl{fig01}) $J_{ij} = -1$ spin-spin interactions. (b)
Possible combinations of the spin-spin interactions to a vertex of model
system $C_1$. No zero-energy spins are possible if all neighboring spins
point in the same direction.}
\end{figure}

\begin{figure}
\includegraphics[width=\linewidth]{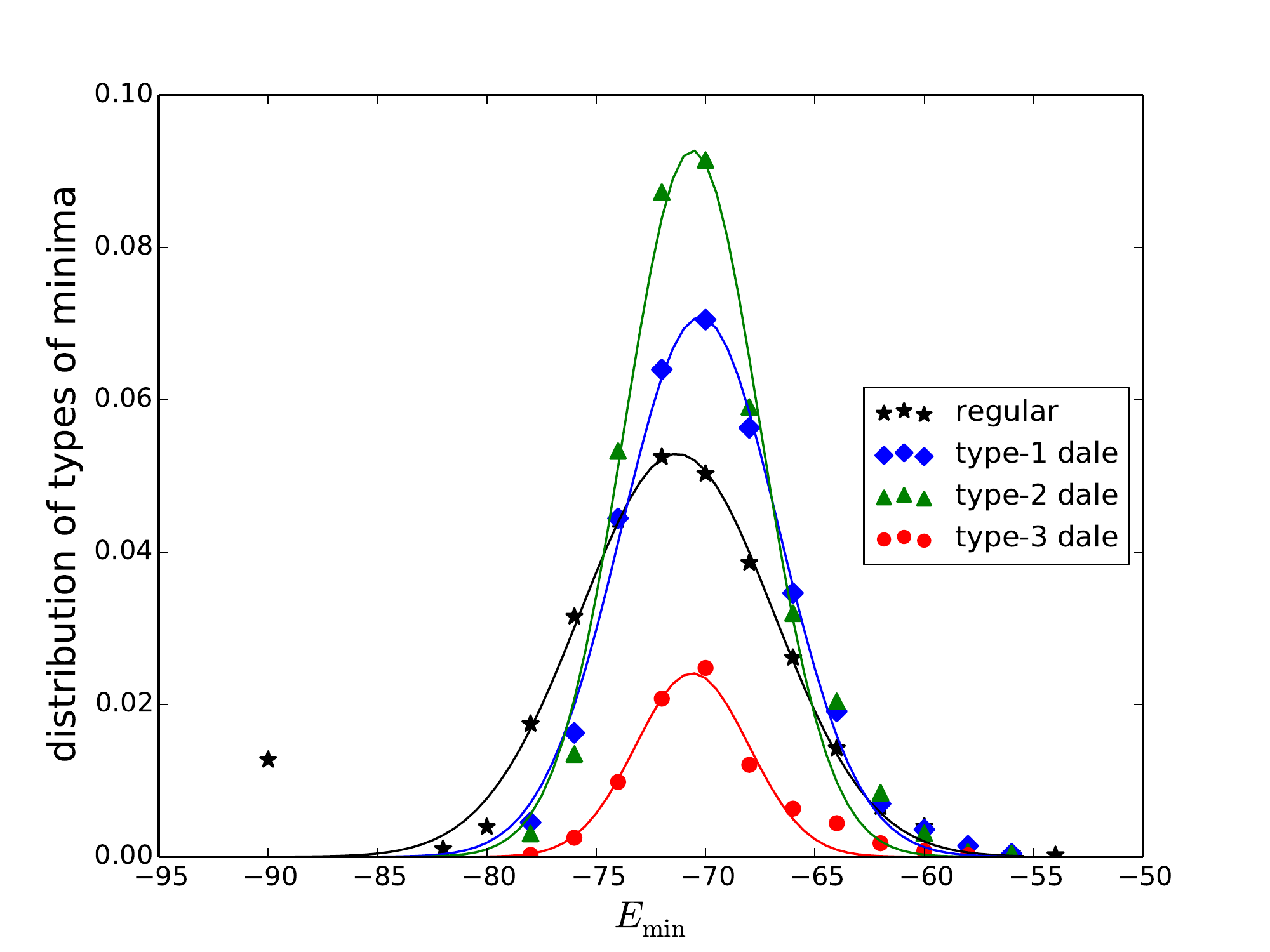}
\caption{\label{dis_pseu_p1_1}
Distribution of the different dale minima types for the $C_1$ model
system of planted spin glasses. The lines are Gaussian fits and guides
to the eye
\cite{comment:daledist}.}
\end{figure}

Figures~\ref{dis_graph_p1_1}, \ref{dis_graph_p2_1} and
\ref{dis_graph_p3_1} show examples of the disconnectivity graphs of the
$C_1$, $C_2$, and the $C_3$ model systems, respectively.  As can be seen
from Fig.~\ref{dis_graph_p1_1} (see also Fig.~\ref{dis_pseu_p1_1}), the
ground state of the $C_1$ model system has an energy of $E_{\rm ground}
= -90$. This is due to the fact that each unit cell has an energy of
$H=-5$ corresponding to a configuration in which either all spins are up
or all spins are down and there are $18$ unit cells in the lattice studied.
Furthermore, the ground state of the $C_1$ model system is always a
regular minimum.  Figure~\ref{C1_planted}(a) shows the possible
combinations of spin-spin interactions to a vertex belonging to a single
unit cell. Each unit cell can contribute to a vertex either two
$J_{ij}=+2$ or one $J_{ij}=+2$ and one $J_{ij} = -1$ spin-spin
interaction. Since two unit cells join at each vertex, this gives the
possible combinations of spin-spin interactions shown in
Fig.~\ref{C1_planted}(b). As can be seen, none of the possible
combinations of spin-spin interactions leads to a zero-energy vertex if
all neighboring spins have the same orientation. Hence, the ground
states of the $C_1$ model are always regular minima.

\begin{figure}
\includegraphics[width=\linewidth]{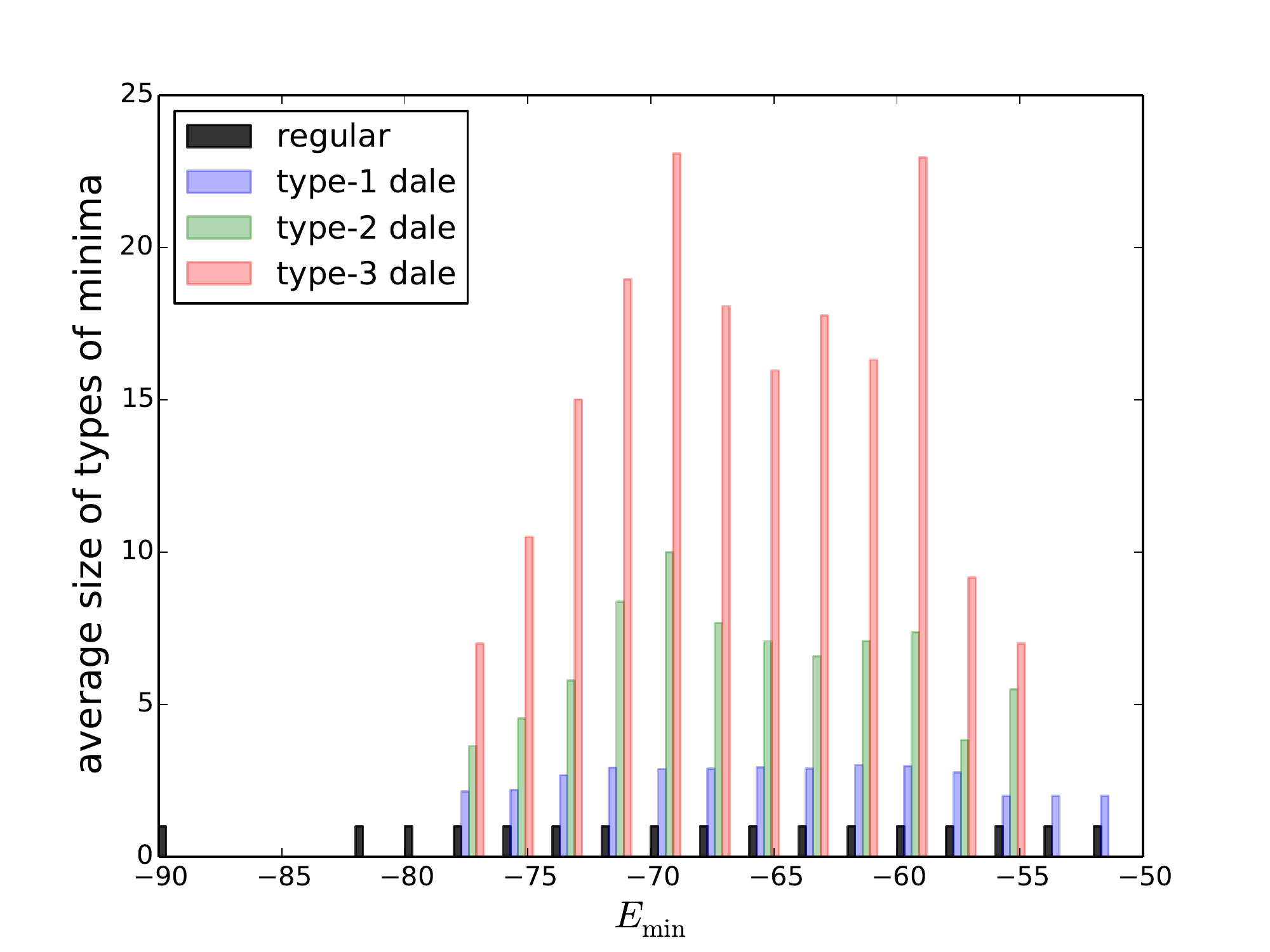}
\caption{\label{size_pseu_p1_1} Average size of the different dale 
minima types for the $C_1$ model of planted spin glasses.}
\end{figure}

Figure~\ref{dis_pseu_p1_1} shows the distribution of the minima types of
the $C_1$ model system for all the minima.  One can see in
Fig.~\ref{dis_pseu_p1_1} that the ground states of this model are always
regular minima and are separated by a large energy gap to higher energy
states.  It appears that the $C_1$ model system has on average the
highest number of states belonging to type-2 dale minima 
(see Tab.~\ref{table1}). This suggests
that with a simple optimization routine one is most likely to encounter
type-2 dale minima lying in the median regions of the energetically
possible configurations.

Figure~\ref{size_pseu_p1_1} shows the average size of the minima and
types of dale minima at their respective energies. With size we
understand here the number of subminima that make up the different types
of dales. Regular minima consist of only one subminimum, hence their
size is always one at any energy level on which they occur. The size of
the different types of dales is always larger than that of the regular
minimum, with type-3 dales having the highest number of subminima
because these are composed of two dales of type-1 or type-2 joined
together. The occurrence that type-1 dale minima are smaller than type-2
dale minima at all energy levels (Fig.~\ref{size_pseu_p1_1}) is not a
general property of the types of dales. It is a specific feature of the
model studied. Furthermore, as can be seen from
Fig.~\ref{size_pseu_p1_1}, the number of subminima belonging to the
individual dales is especially large around medium energies (at about
$E_{\mathrm{min}}=-70$) with a second peak at $E_{\mathrm{min}}=-60$.
Taking into account the findings of the distribution of the types of
minima (Fig.~\ref{dis_pseu_p1_1}) only the first peak considerably
contributes to the features of the energy landscape. The second peak,
while comparably large in the size of type-2 and type-3 dale minima, has
a low occurrence (less than $0.01$, see Fig.~\ref{dis_pseu_p1_1}) and
thus has little significance for general optimization procedures.  The
size of type-1 dale minima does not vary much and is always less than
four for the system sizes studied. 

\begin{figure}
\includegraphics[width=\linewidth]{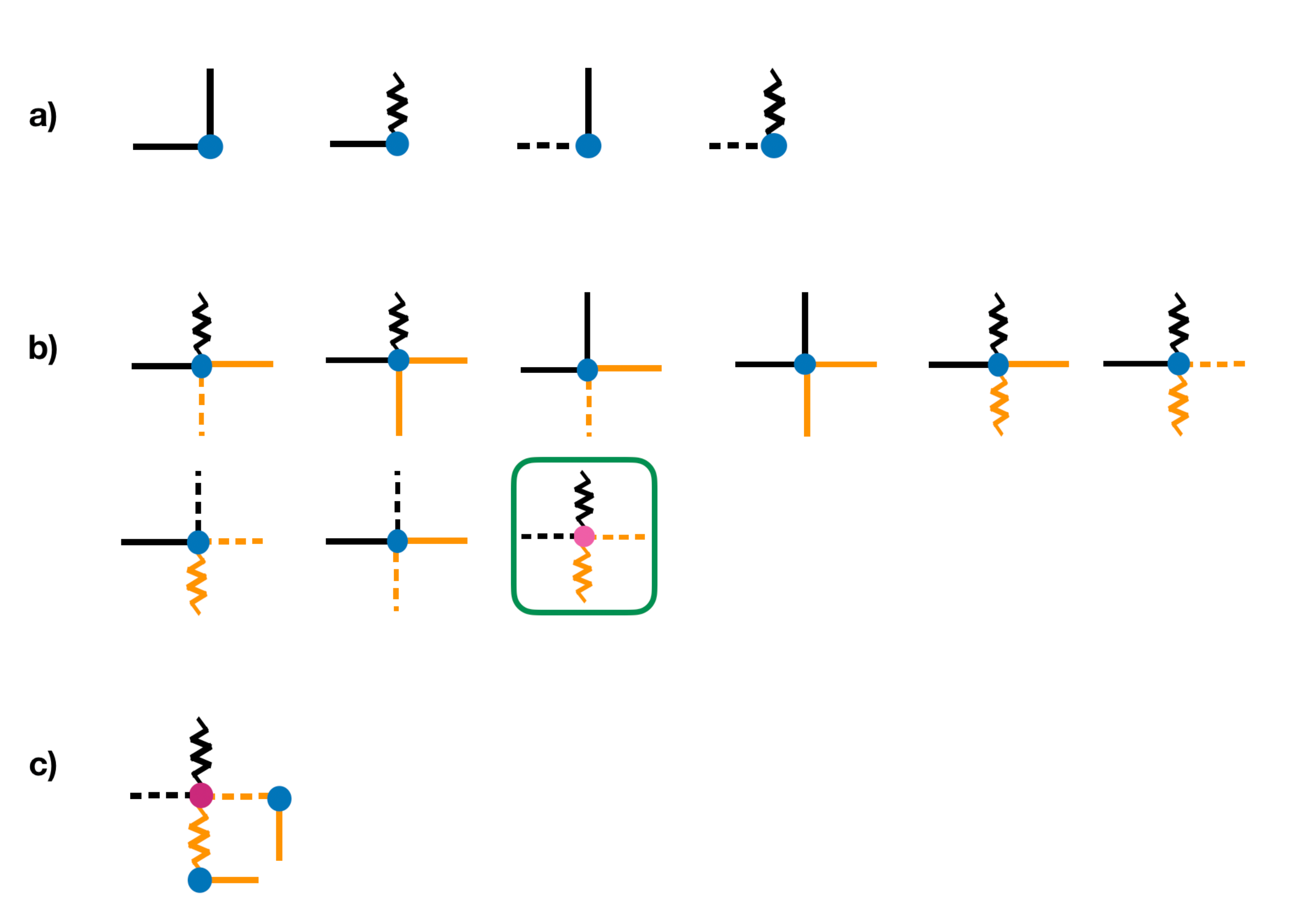}
\caption{\label{C2planted} 
(a) Possible contributions of spin-spin interactions of a single $C_2$
unit cell to a vertex. The bonds $J_{ij} = -1$, $1$, and $2$ are
indicated by wiggly lines (\protect\imgl{fig01}), dashed lines
(\protect\imgl{fig02}), and solid lines (\protect\imgl{fig03})
respectively. (b) Possible combinations of the spin-spin interactions to
a vertex of model system $C_2$. Only one of the combinations (framed)
allows for zero-energy spins for the planted solutions. (c) The
zero-energy vertex of case (b) only leaves $J_{ij}=+2$ spin-spin
interactions for each joining unit cell, hence this system cannot have
neighboring zero-energy spins in the planted solution. }
\end{figure}

The ground states of the $C_2$ model are either regular minima or type-1
dale minima and have a ground-state energy of $E_{\rm ground}=-72$,
because each of the $18$ cells has an energy of $H=-4$.  For the
planted ground-state solutions the occurrence of type-1 dale minima can
be understood by considering the possible combinations of spin-spin
interactions to the spins. Figure~\ref{C2planted} illustrates this
situation. The possible contributions of spin-spin interactions of a
single unit cell to a vertex are shown in Fig.~\ref{C2planted}(a). In
Fig.~\ref{C2planted}(b) the possible distinct combinations of spin-spin
interactions to a vertex are shown. Only one of these combinations
allows for a zero-energy spin in the planted solutions, i.e., two unit
cells need to be oriented in such a way, that from each of the cells one
$J_{ij}=+1$ and one $J_{ij}=-1$ spin-spin interaction join at the
vertex. Only then the energy of the vertex will be balanced if all
neighboring spins point in the same direction. However, because each of
the two unit cells already contributes with their $J_{ij}=+1$ and
$J_{ij}=-1$ spin-spin interactions, the remaining spin-spin interactions
of each of the two unit cells are $J_{ij}=+2$ as illustrated in
Fig.~\ref{C2planted}. This makes it impossible for any of the
neighboring vertices to the zero-energy vertex to be energetically
balanced in the planted solution.  This prohibits the formation of
connected zero-energy spins, and hence the planted solutions of this
model will always be regular minima or type-1 dale minima.

\begin{figure}
\includegraphics[width=\linewidth]{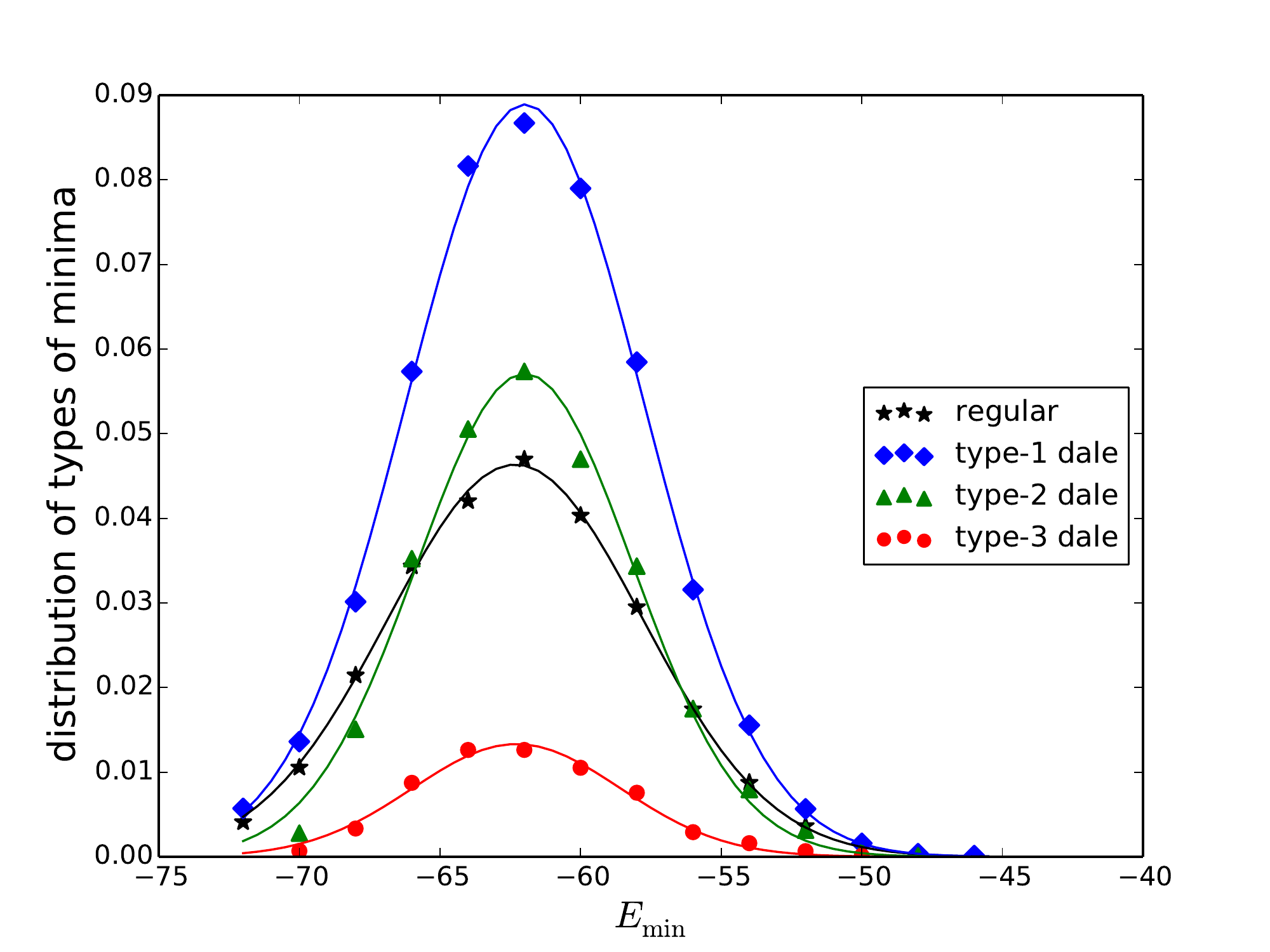}
\caption{\label{dis_pseu_p2_1}
Distribution of the different dale type minima for the $C_2$ model of
planted spin glasses. The lines are Gaussian fits and guides to the eye
\cite{comment:daledist}.}
\end{figure}

\begin{figure}
\includegraphics[width=\linewidth]{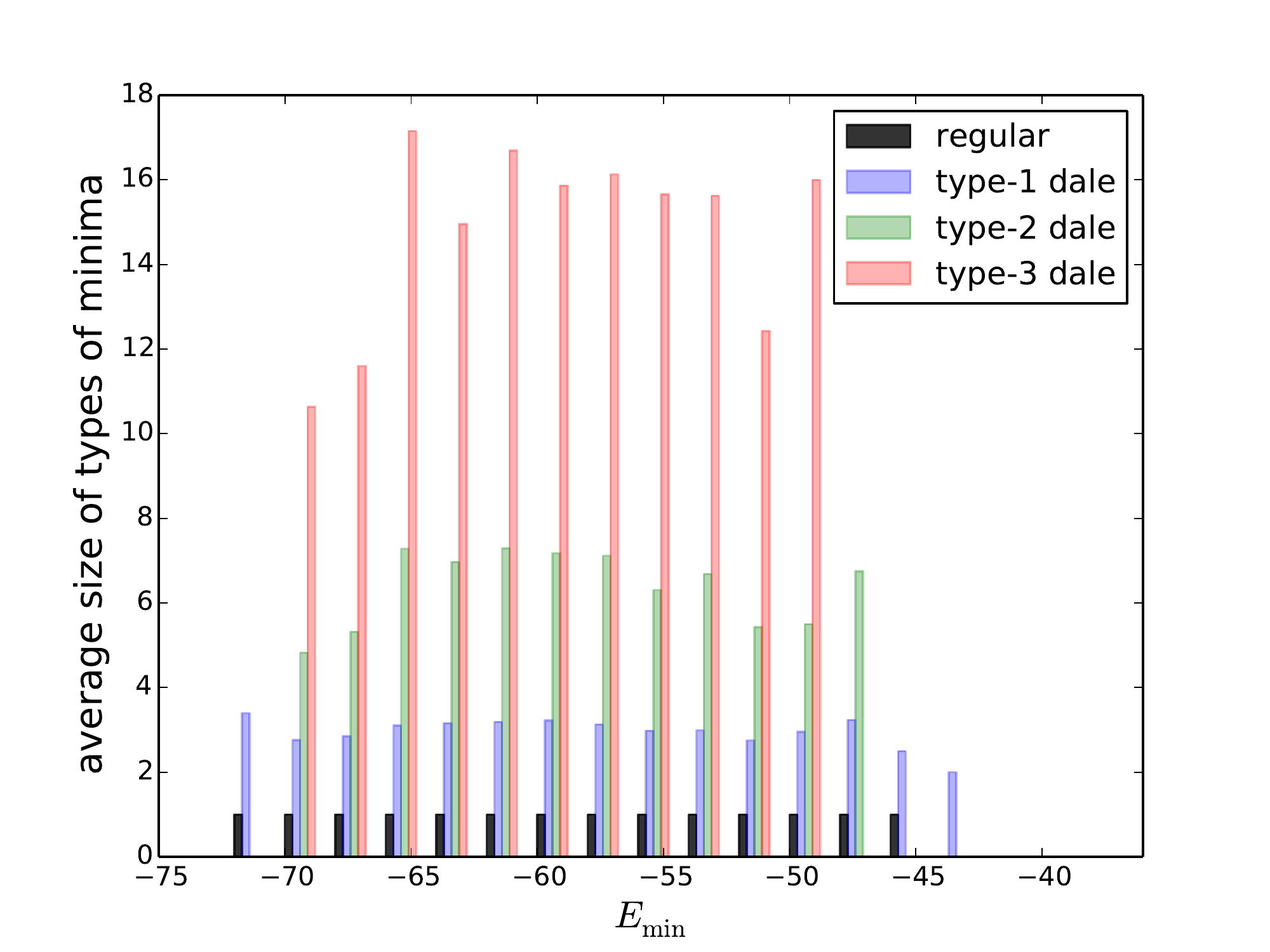}
\caption{\label{size_pseu_p2_1} Average size of the different dale minima 
types for the $C_2$ model of planted spin glasses.}
\end{figure}

Figure~\ref{dis_pseu_p2_1} shows the distribution of minima types in the
reduced description of all the minima in the $C_2$ model.  Different to
the $C_1$ model system, the $C_2$ model does not have a large gap to the
ground state.  Similarly to $C_1$, the majority of the minima populate
the energetically medium regions. Together with the higher number of
minima, this suggests that this type will be harder to solve in
optimization routines than $C_1$ and $C_3$. This was also observed in
the study of Perera {\em et al.}~\cite{perera:19}, which compares the
time to solution of different models of tile-planted spin glasses.

Figure~\ref{size_pseu_p2_1} shows the average size of the types of
minima at their respective energies. The size of types of minima follows
the same order as in the $C_1$ model, with regular minima always having
size one, followed by type-1 dale minima, type-2 dale minima, and type-3
dale minima having the largest size. In contrast to $C_1$ in the $C_2$
model, no distinct peak in the size is visible. The largest sizes of the
dales seem to span a range from $E_{\mathrm{min}}=-66$ to
$E_{\mathrm{min}}=-54$ for the system size studied. However, taking into
account the high occurrence of dales at the energy of
$E_{\mathrm{min}}=-62$ (see Fig.~\ref{dis_pseu_p2_1}) indicates that
especially energies in this range have on average the highest number of
states pertaining to minimum energy configurations.

\begin{figure}
\includegraphics[width=\linewidth]{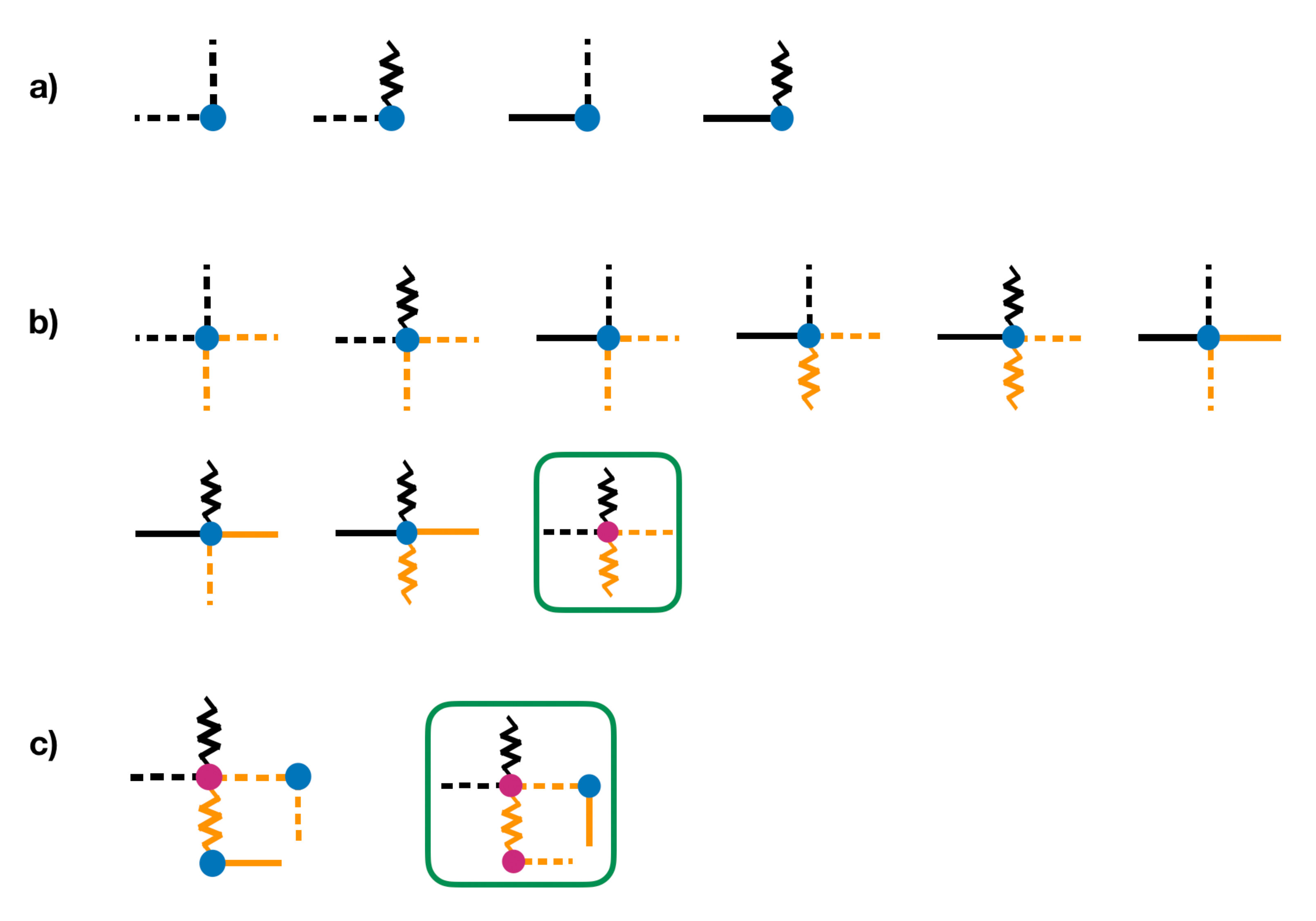}
\caption{\label{C3planted} 
(a) Possible contributions of spin-spin interactions of a single $C_3$
unit cell to a vertex. The bonds $J_{ij} = -1$ ,$1$, and $2$ are
indicated by wiggly lines (\protect\imgl{fig01}), dashed lines
(\protect\imgl{fig02}) and solid lines (\protect\imgl{fig03}),
respectively. (b) Possible combinations of the spin-spin interactions to
a vertex of model system $C_3$. Only one of the combinations (framed)
allows for zero-energy spins of the planted solutions. (c) Possible
combinations of the spin-spin interactions for the unit cells joined at
a zero-energy vertex. One of these combinations (framed) allows for a
zero-energy spin at a second vertex and hence for the formation of
connected structures of zero-energy spins.}
\end{figure}

Figure~\ref{dis_graph_p3_1} shows an example of the disconnectivity
graph for the $C_3$ model.  Due to the small values of the spin-spin
interactions, the ground state of the $C_3$ model is highly degenerate.
It exhibits a large number of minima of all types and has the energy
$E_{\rm ground}=-54$ as each of the $18$ cells has an energy of $H=-3$.  Figure
\ref{C3planted} (a) shows the possible contributions of the spin-spin
interactions of a single unit cell to a vertex. If two unit cells are
joined at a vertex, the possible distinct combinations of spin-spin
interactions to a vertex are shown in Fig.~\ref{C3planted}.  Note that
from these combinations, similar to the $C_2$ model system, only one
configuration of spin-spin interactions leads to a zero-energy spin.
This combination is framed in green in the figure. However, different
from the $C_2$ case, it allows for larger structures of connected
zero-energy spins (see Fig.~\ref{C3planted}). These connected structures
allow for type-2 dale minima and since the spin-spin interactions to
spins that are connected to zero-energy spins can be balanced once the
zero-energy spins have been flipped, this allows also for type-3 dale
minima to occur. Hence, the planted solutions of model system $C_3$ can
be regular minima or dale-minima of any type.

\begin{figure}
\includegraphics[width=\linewidth]{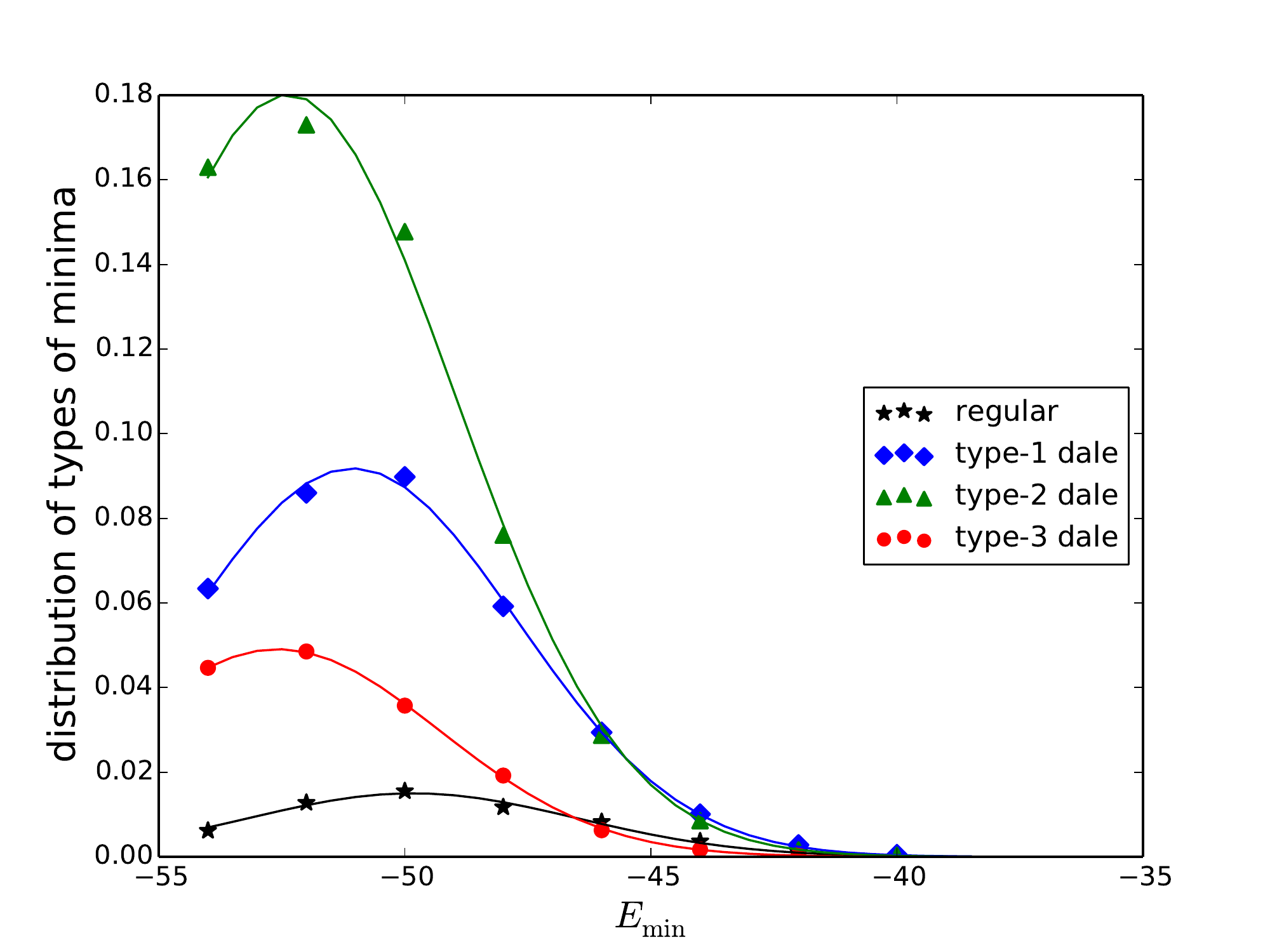}
\caption{\label{dis_pseu_p3_1}
Distribution of the different dale types for the $C_3$ model system of
planted spin glasses. The lines are Gaussian fits and guides to the
eye
\cite{comment:daledist}.}
\end{figure}

\begin{figure}
\includegraphics[width=\linewidth]{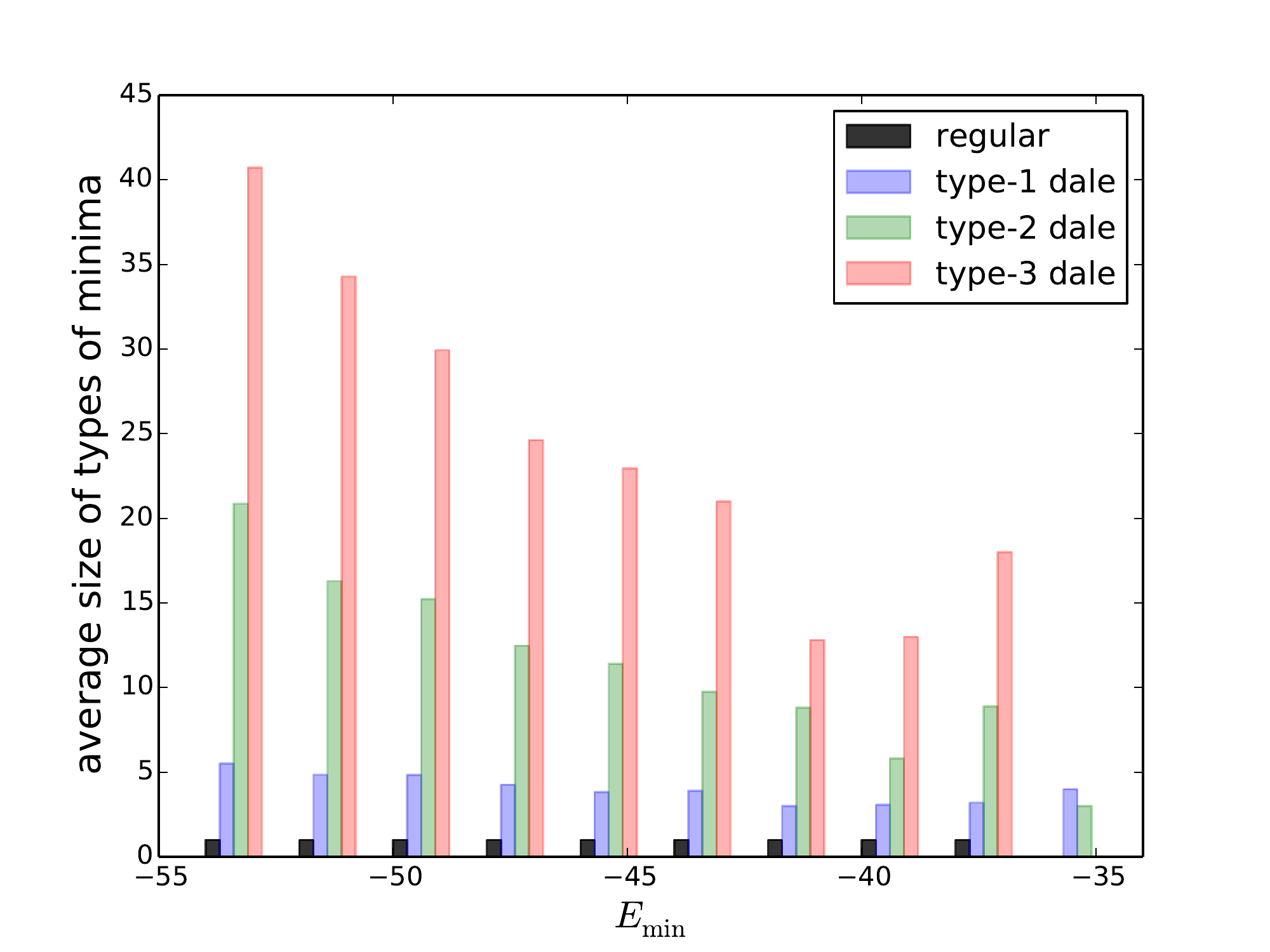}
\caption{\label{size_pseu_p3_1}Average size of the different dale types for 
the $C_3$ model system of planted spin glasses. }
\end{figure}

Figure~\ref{dis_pseu_p3_1} shows the distribution of the minima types in
the reduced description of the $C_3$ model system. Different to the
$C_1$ and $C_2$ model systems, a large number of minima is energetically
close to the ground state. This model is dominated by type-2 dale minima
suggesting the formation of large minimum-energy dales on the energy
landscape, i.e., on average each dale consists of $17$ individual
configurations connected by zero-energy spins (see Tab.~\ref{table1}).
Furthermore, this model has the least number of regular minima at any
energy. These results suggest that, using standard optimization
procedures, a configuration belonging to the ground state will be easier
to find for the $C_3$ model than for the other models. This is in
agreement with the numerical results of Perera {\em et al.}
\cite{perera:19}.

Figure~\ref{size_pseu_p3_1} shows the average size of the types of
minima at their respective energies of the $C_3$ model. As can be seen,
similar to the $C_1$ and $C_2$ model, regular minima always have size
one, i.e., they consist of only one subminimum, and type-3 dale minima
have the largest number of subminima. However, different from both the
$C_1$ and $C_2$ models, type-2 dale minima are not always larger than
type-1, i.e., only for energies up to $E_{\mathrm{min}}=-38$ the type-2
dale minima consist of more subminima than the type-1 dale minima, but
at the highest occupied energy level at $E_{\mathrm{min}}=-36$ type-1
dale minima are on average larger than type-2 dales. The size of type-1
dale minima is relatively constant ($\leq 5$) for all energy levels.
Type-2 and type-3 dales are largest at the lowest energies, which
together with their very high occurrence (compare to
Fig.~\ref{dis_pseu_p3_1}) is a further indication of the relative ease
at which lowest-energy states can be found.

\section{Summary}

We introduce a classification scheme for the different types of
connected degenerate minima, which we call dale minima.  We distinguish
between regular minima, i.e., minima for which a flip of any spin would
increase the energy, and dale minima.  The dale minima form broad
valleys in the energy landscape and are composed of multiple minima
connected via zero-energy spin flips. We distinguish between three
different types of dale minima based on similarity and accessibility of
the states to each other. This procedure effectively helps to reduce the
number of stored minima during the computation and eases the
visualization of disconnectivity graphs. In the disconnectivity graphs,
the dale minima are distinguished by colors. Furthermore, we add a bar
chart depicting the average number of subminima pertaining to the dales
at the respective energy levels. The added classification into different
types of minima allows for an enhanced version of disconnectivity graphs
that give an intuitive understanding of the important aspects of the
potential energy landscape of spin systems.

We apply our classification scheme to a spin-glass model with planted
solutions and differentiate between the three elementary problem
classes, namely $C_1$, $C_2$, and $C_3$. The resulting disconnectivity
graphs and subsequent analysis show distinctly different features for
the different classes. $C_1$ systems only have two ground states,
corresponding to the ferromagnetic solution of all spins up or all spins
down. Its higher-energy states are separated by a large gap to the
ground state and are highly degenerate. All of the states, including the
ground states, of the $C_2$ and $C_3$ model systems are degenerate, but
do not have a large gap to the ground-state. Compared to $C_1$ and
$C_2$, $C_3$ instances have the highest number of ground-states relative
to higher energy states, leading us to conclude that $C_3$ instances are
the easiest to solve when using conventional optimization routines.
Plots of the different minima types versus their energy show distinct
Gaussian features, with the mean of the $C_1$ and $C_2$ models being
centered at energies above the ground state. Only for $C_3$ instances
are the majority of the minima located close to the ground state.

While our analysis of these types of planted spin glasses is limited to
a system size of $36$ spins, we estimate that the general features of
the energy landscape, such as the occurrence of ground-state solutions
and the relative distribution of the minima across different energy
levels, is preserved for larger system sizes. Similar arguments hold
for the occurrence of the different types of dales and their
distribution over the different energy levels. Note, however, that our
conclusions are based on relatively small system sizes because the phase
space grows exponentially with the number of spins.  This means that
further types of dales might occur when the system sizes are increased.
However, the developed methods can be generally applied, thus presenting
a simple visual approach to characterize the phase space of statistical
mechanical systems.

\begin{acknowledgments}

We would like to thank Dilina Perera for feedback on the manuscript. 
H.G.K.~would like to thank Izumi Kirkland for inspiration.  This
research is based upon work supported in part by the Office of the
Director of National Intelligence (ODNI), Intelligence Advanced Research
Projects Activity (IARPA), via MIT Lincoln Laboratory Air Force Contract
No.~FA8721-05-C-0002.  The views and conclusions contained herein are
those of the authors and should not be interpreted as necessarily
representing the official policies or endorsements, either expressed or
implied, of ODNI, IARPA, or the U.S. Government.  The U.S. Government is
authorized to reproduce and distribute reprints for Governmental purpose
notwithstanding any copyright annotation thereon.  We thank the Texas
A\&M University for providing high performance computing resources.

\end{acknowledgments}

\bibliography{refs,comments}

\begin{thebibliography}{49}
\bibitem[{com({\natexlab{a}})}]{comment:name}
\bibinfo{note}{These dales are sometimes also referred to as U-shaped minima
  \cite{garstecki:99}, clusters
  \cite{hartmann:00a,hartmann:00b,krawczyk:02,hartmann:02,mann:10}, or local
  ensembles of ground states.}
\bibitem[{com({\natexlab{b}})}]{comment:hi}
\bibinfo{note}{We use the word dale to highlight the discrete nature of the
  energies of our systems.}
\bibitem[{com({\natexlab{d}})}]{comment:structure}
\bibinfo{note}{One would expect the overall structure of the cluster to be
  highly symmetric due to the symmetry regarding a flip of all spins of the
  individual minima. This, however, is a feature of the highly
  multi-dimensional nature of the energy landscape and can only be hinted at in
  two-dimensional representations.}
\bibitem[{com({\natexlab{e}})}]{comment:table}
\bibinfo{note}{Note, that since type-3 dale minima consist of type-1 and type-2
  dale minima, which are joined by directional paths, the values given in this
  column relate only to the total number of minima, but not to the other types.
  They are to be interpreted separately.}
\bibitem[{com({\natexlab{f}})}]{comment:daledist}
\bibinfo{note}{Note that type-3 dales consist of type-1 and type-2 dales joined
  together. In order to avoid double counting, this figure is normalized to
  unity taking into account only regular minima, type-1, and type-2 dales. The
  values for type-3 dales are with respect to the total number of minima of the
  systems.}
\end{thebibliography}

\end{document}